# Longons from the nonlinear dispersion of Galerkin regularization

Jian-Zhou Zhu (朱建州)[1]

[1]Su-Cheng Centre for Fundamental and Interdisciplinary Sciences, Gaochun, Nanjing 211316, China

Irregular compactons and peakons from some nonlinear dispersions can be regularized by another type of nonlinear dispersion, defined by a pseudo-differential operator in physical space for the Galerkin truncation preserving finite Fourier modes of wavenumbers no larger than $K$. This resembles yet differs from the Korteweg-de Vries (KdV) regularization of the Burgers-Hopf (BH) equation. The Galerkin-regularized compacton, peakon, KdV, and BH dynamics exhibit novel traveling waves and interacting solitonic structures amidst weaker, less-ordered components ('longons'). Quasi-periodic solutions are also constructed with on-torus invariants, towards a potential Kolmogorov-Arnold-Moser (KAM) argument with presumably whiskered tori. The latter are persistent against the approximation of the truncation by a type of linear dispersion models, resulting in similar longulent states. Time-dependent and stationary behaviors in the large-$K$ limit are addressed with numerical results.

## I. INTRODUCTION

Any physical theory, including even the standard model of particle physics, can be viewed as an effective theory truncated from a more comprehensive one [1]. Determining when and how this truncation or regularization, which often may be characterized by the pseudo-differential operator $\Pi$ defined through a cutoff/symbol function $p$, influences the outcomes and predictions is generally difficult, as it is often non-trivial to formulate a clear dynamic description of the truncation within an appropriate theoretical framework.

Relatively simple are likely those systems that would otherwise yield singular or irregular objects, as the regularization such as that retaining finite Fourier modes (Galerkin truncation/regularization) of wavenumber no larger than $K$, say, shall smooth the latter, apparently. However, determining the precise regularization effects is challenging, even for the simplest classical systems. For instance, could new '(quasi-)particles' or, correspondingly, in classical physics, novel coherent structures, such as solitons, or, unexpected phenomena emerge and, if so, what are their long-time behaviors? Furthermore, the truncation effects may severely depend on the situations, like the classical integrable systems: as Chen and Olver remarked [2], Zabusky and Kruskal [3] could have discovered the quantum revival and fractalization instead of the soliton, had they used the piecewise-constant rather than sinusoidal initial data.

Furthermore, if the regularization is formally reduced to be vanishingly small, does it induce anomalous effects? A prominent example is the Fermi-Pasta(-Tsingou)-Ulam ('FPU') recurrent problem [4] and the associated Korteweg-de Vries (KdV) soliton [3], which originally inspired the speculation for similar effects from Fourier Galerkin truncation on general hydrodynamical-type systems, including the high-dimensional ones [5] that are left to be further examined [6]: particularly clear in the KdV system now, the linearly dispersive regularization, not as sharp as the Galerkin regularization (Gr) though, of the Burgers-Hopf (BH) equations can lead to "solitization" [3, 7] and to quantum rivival and fractalization [2, 8, 9]; so, anti-thermalization results, not as Lee [5] and others postulated in different situations [10], appear not impossible for general conservative regularizations [including the hyper-KdV (hKdV, to be introduced below) linear dispersion], at least in one-component dynamics such as the compacton and peakon (CP) [11–16], besides BH, systems. And, the well-known zero-dispersion limit issue, naturally leads to ponderations on the $K \to \infty$ limit.

Indeed, it can be shown that the Gr-systems (GrKdV, GrCP, and GrBH etc.) support travelling waves and interacting solitonic structures of clear periodic nature, the latter always being accompanied by weaker, less-ordered components (both called 'longons', with the final explanation of the term given in Sec. IV A), which are absent or differ from those in the corresponding original systems. A shorter account is given in Ref. [17], while more extensive investigation with the help of numerics and further expansion of the idea of torus-specific invariant(s) leading to quasi-periodic tori, associated to the Kolmogorov-Arnold-Moser (KAM) argument, is completed here. This is somewhat reminiscent of the intermediate phase in chaos-integrability transition of, for instance, the standard map and quantum spin chain, and the "longulent" state with disordered structures may be called "longulence". It also appears appropriate to unify the notions of thermalization/chaotization and strong interactions of longons with severe phase-shifts [17], so that terminologies such as (statistical absolute) equilibrium can also apply.

In physics, the Galerkin truncation or regularization has been used to study the effect of thermalization, remarkably the inference of turbulence dynamics as a paradigm following the above mentioned Lee's work. And, in the partial differential equation theory, the Galerkin method has been used for numerical approximation and proof of the existence of solutions. Here, we open a new nonlinear-science direction of investigation: its possibility of leading to (pseudo-)periodic patterns, in particular solitonic structures, which is apparently intimate to the soliton phenomena and theories of integra-bility, invariant tori, persistence and KAM etc. In particular, it is possible to rigorously prove the persistent or KAM whiskered tori associated to the longulent states in the a-posterio fashion developed recently (c.f., Ref. [18] on partial differential equation and Ref. [19] on maps, and references therein).

The above mentioned connection with integrable theory requires a brief background of the latter as follows. The KdV hyperbolic secant/$\mathrm{sech}^2$ 'solitary wave' solution had been used prior to Zabusky and Kruskal's discovery of the 'solitons' with periodic boundary condition for which the cnoidal/$\mathrm{cn}^2$ travelling-wave solution applies. The periodic KdV (multi-)soltion-analogue problem were first solved around 1975 by various independent groups [20, 21] [see, e.g., Osborne [22] for solutions of also other periodic systems and their applications, including of course other models such as the nonlinear Schrödinger (NLS) equation], with also subsequent infinite-gap results which we do not have a clear correspondence for lack of the operator in the Lax pair, at least at this point. For the results here in the domain of period $L$, the large-$L$ behaviors appear consistent. Considering the less mathematically rigorous terminologies like 'incoherent' and 'dissipative' solitons, we tend to adopt the usage of 'solitary wave' and 'soliton' for periodic boundary conditions as well, for simplicity. Actually, we will see that the center manifold theory and homoclinic orbit argument used in Ref. [15] for solitary waves extend to the hKdV model approximating and converging to the truncation. As noted in Ref. [17], the number of various traveling-wave solutions increases rapidly with $K$ in each Gr-system, but only that whose profile resembles the cnoidal wave (usually with however wiggles in between the strong pulses) of a single peak seems to present as the shape of the apparently solitonic longon(s) in the well-developed state. On the other hand, various longon profiles do have been observed in the transient states or less-ordered (chaotic-looking) longons. For general traveling-wave solutions of Gr-systems, the term 'solitary wave' should be used with caveat if strict mathematical rigor is required. Also, interacting solitonic longons, always accompanied by weaker less-ordered ones (reminiscent of the strange particles and resonances observed among strongly interacting particles), in Gr-systems do appear to be different to conventional solitons. This motivated the association to whiskered tori and on-torus invariants for augmenting a KAM argument [17], here further discuss this point with also numerical computations using the data for the exact quasi-periodic solutions specified by the additional torus-specific functionals.

Let us take a step back to the beginning with the CP model [13, 14]

$$v_t \pm v_{txx} + 3vv_x = \mp 2v_xv_{xx} \mp vv_{xxx} \tag{1}$$

where the upper signs are for compacton and lower for peakon. Note that the compacton branch can be viewed as the (integrable) extension of the Rosenau-Hyman (RH [11]) $K(2,2)$ model, while the peakon branch is actually the Camassa-Holm model (CH [12]) whose signs Eq. (1) follows. The discrepancy to the signs in Ref. [14] disappears with a mere change of coordinate $x \mapsto -x$. Li and Olver [15, 16] used the system,

$$v_t + \gamma v_{txx} = \alpha v_x + \beta v_{xxx} + \frac{3}{\gamma} vv_x + 2v_xv_{xx} + vv_{xxx}, \tag{2}$$

to study the solitary waves and their convergence to the more general parameterized compacton/peakon, which appears relevant to the linearly dispersive model approximation below for the Galerkin regularization.

Olver and Rosenau [14] already noted that "the complex transformation $x \mapsto \hat{i}x$ and $t \mapsto \hat{i}t$ [with $\hat{i}^2 = -1$] will interchange ..., indicating a close interconnection between compactons and peakons." Such an observation, to our point of view, may imply that the two CP branches unify somewhere in the complex plane to which they are continued. The next question is about the large-$K$ limit. Since the truncation is an essentially more severe regularization than the nonlinear dispersion in CP, GrCP should not reduce to the latter. A systematic investigation of this problem is beyond the purpose here, but the answer for the same question of GrBH actually was already offered [23, 24]. And, the problem resembles the zero-dispersion limit of the Korteweg-de Vries equation which has been well established in the well-known Lax-Levermore work (Venakides [25] for the periodic case), which leads to the introduction of the following models. [Another purpose of this note is also to try to bring such mathematically intricate and involved issues to a more physically intuitive level, with the hope that the discussions can be more motivative and illuminating in some way.]

Let $v(x,t)$ solve, with $x$-period $2\pi$ and $v_0 = v(x,0)$,

$$v_t + vv_x = a. \tag{3}$$

$a = 0$, $\mu v_{xxx}$, and $\nu(v^2)_{xxx}$ identify, respectively, the BH, KdV and RH [or $K(2,2)$] equations. For analytical $v$, of course KdV and RH reduce to BH with vanishing $\mu$ and $\nu$ respectively, which however is not the case for nonanalytical $v$. The CP model presented earlier can also be rescaled into this form.

We then have $\hat{v}_k = \int_0^{2\pi} \frac{v}{2\pi} e^{-\hat{i}kx} dx$, with complex conjugacy ($c.c.$) $\hat{v}_k^* = \hat{v}_{-k}$ for reality. Additionally, $v\partial_x v = \sum_k \hat{b}_k e^{\hat{i}kx}$ where $\hat{b}_k = \frac{\hat{i}k}{2} \sum_p \hat{v}_p \hat{v}_{k-p}$ and $\hat{i}^2 = -1$. For $v_0$ well-prepared in ${}^K\mathbb{G} = \{k : -K \le k \le K\}$ ("Galerkin space" hereafter), we can calculate each $\hat{b}_m$ for $K < |m|$ ($\le 2K$). In the BH case, setting $\hat{a}_m$ to be ${}^K\hat{g}_m = \hat{b}_m$ for $m \notin {}^K\mathbb{G}$ and 0 otherwise results in Galerkin truncation: for all $m \notin {}^K\mathbb{G}$, $\hat{v}_m(t) \equiv 0$ ($t > 0$), thus the GrBH; and, similarly, the other GrKdV, GrCP, or, GrCH and GrRH systems with their respective ${}^Kg$.

In the current periodic setup, a pseudo-differential operator $\Pi$ on $v$ can be defined through its symbol $p(x,k)$:

$$\Pi[v](x) = \sum_k e^{\hat{i}kx} p(x,k) \hat{v}_k. \tag{4}$$

Such formal mathematical objects are popularly used in signal processing and physics, associated to the practical notions of 'cutoff' and 'filtering'.

Various $p$s for physically favorable cutoffs may be introduced into the corresponding Lagrangian and subsequent regularized dynamical equations as effective field theories of, say, condensed matter and particle physics. Such ideas and techniques are of course widely applicable in many fields, and the reason we particularly refer to such effective field theories is because they involve '(quasi-)particles' which are related to the discovery of emergence of new "particles" (to be called 'longons') from the following pseudo-diferential operator $P_K$ with $p$ being the indicator function of ${}^K\mathbb{G}$ for a 'hard cutoff'. [And, as will be shown further below, the effective models 'softening' such a hard cutoff also present solitonic structures close to the longons, which means some generality of the results among effective field theories applying different cutoffs.]

Define $P_K v(x) := \sum_{|k|\leq K} \hat{v}_k \exp\{\hat{i}kx\} =: u$, $B := u^2/2$ and ${}^KG := B - P_K B$. Then follows the GrBH equation [23]

$$Du/Dt := \partial_t u + \partial_x B = \partial_x {}^KG;\ u_0 = P_K v_0. \tag{5}$$

Or, in $k$-space, $\hat{u}_k$ satisfies

$$\partial_t \hat{u}_k^\# = -\frac{\hat{i}k}{2} \sum_{p,q,p+q=k\in{}^K\mathbb{G}} \hat{u}_p^\# \hat{u}_q^\#. \tag{6}$$

The Galerkin force ${}^Kg = \partial_x {}^KG$, with ${}^K\hat{g}_m$ for $K < |m| \leq 2K$, is excited when there exists $\hat{u}_k \neq 0$ with $k > K/2$ and can develop to $O(b)$ with $b = \partial_x B$ [or $O({}^Kb)$ with ${}^Kb = b - {}^Kg$]. It should persist with $g = \lim_{K\to\infty} {}^Kg$, formally (for possibly being infinite or nonconvergent): it is intuitively clear and possible to be made rigorous in the language of nonstandard analysis [26] that $g$ acts on the wavenumbers beyond the infinite $\pm K$ to excite microscopic longons at infinitesimal scales. Unlike the well-known zero-dispersion KdV limit, there is no external control parameter for ${}^Kg$ or $g$. It is envisaged that the potential singularity to be developed become its own revenge with $g$ excited to balance $b$ and effecting no longon dissipation (unlike the Burgers shock with vanishing viscosity).

Other Gr-systems, such as Eq. (10) for GrCP below, can be accordingly made precise. [For distinguishment, we could have used notations such as ${}^K_{bh}g = \partial_x({}^K_{bh}G)$ etc., but such self-evident index as '${}_{bh}$' is neglected for simplicity when it is clear in the context as in the above.] The CP Hamiltonian operator $J_{CP} = -2\pi(\partial_x \pm \partial_x^3)$ in Fourier representation still applies with truncation and is inherited by GrCP, just like $J_{KdV} = -2\pi\partial_x$ by GrKdV [27]. The GrBH reduction replaces the reduced Hamiltonian $\mathcal{H}_{BH} = \int_0^{2\pi} \frac{v^3 dx}{12\pi}$ (for analytical $v$) with [28, 29]

$$\mathcal{H} = \sum_{p,q,k=p+q\in{}^K\mathbb{G}} \hat{u}_k^* \hat{u}_p \hat{u}_q/6 = \int_0^{2\pi} \frac{u^3 dx}{12\pi}, \tag{7}$$

thus the Galerkin interaction potential ${}^K\mathcal{G} = \mathcal{H}_{BH} - \mathcal{H}$. [Interestingly, (Gr)RH Hamiltonian function [30] is of the same form as the above for (Gr)BH.] The other reduced Hamiltonian operator $J'_{BH} := -(u\partial_x + \partial_x u)/3$ involves $u$ and is not transferable to GrBH to facilitate the bi- or tri-Hamiltonian machinery [14]. Actually, only three GrBH invariants, $\mathcal{H}$, $\mathcal{E} = \int_0^{2\pi} \frac{u^2 dx}{4\pi}$ and $\mathcal{M} = \hat{u}_0 = \int_0^{2\pi} \frac{u dx}{2\pi}$ are known for general $K$. Note that no futher truncation is needed for computing $\mathcal{H}$ and $\mathcal{E}$. Similarly, for the GrCP situation, with, e.g., $\mathcal{H}_{CP} = \int_0^{2\pi} \frac{v^3 \mp v(\partial_x v)^2}{4\pi} dx$ and, accordingly, $\mathcal{M}_{CP}$ and $\mathcal{E}_{CP}$, and, their truncated versions. By Galilean invariance, $\mathcal{M}$ is taken to be zero or truncated in this study unless otherwise specified, and $K$ effectively denotes the number of available modes with $2K$ degrees of freedom.

The remainder of the article is organized as follows: Sec. II discusses solitary waves, quasi-periodic tori and interacting longons, including an approximation of the truncation with a linearly dispersive model (Sec. II A 2). Section III details further exploration in the GrBH world, particularly the breakdown of piecewise-constant data with large Hamiltonians into a single solitonic longon (among weaker, less-ordered ones) and the large-$K$ limit, in particular the discussion of constructing whiskered tori with torus-specific invariants. Finally, Sec. IV provides additional discussions, reflecting on past work and future prospects.

## II. TRAVELLING WAVES, QUASI-PRIODIC SOLUTIONS AND INTERACTING LONGONS

We investigate the Gr-system traveling-wave solution $u(x,t) = u(\zeta)$ with $\zeta = x - \lambda t$ up to an arbitrary phase, parameterized by $x_0$ (or $\tau_0$) but absorbed by $x$ (or $t$), thus $\hat{u}_k = \hat{U}_k e^{-\hat{i}k\lambda t}$ for $t$-independent $\hat{U}_k$, i.e.,

$$u_t = -\lambda u_x;\ \partial_t \hat{u}_k = -\hat{i}\lambda k \hat{u}_k. \tag{8}$$

Note that, in some sense, the term "nonlinear wave" can be confusing because the traveling wave described above is clearly linear. The above equation indeed 'formally' linearizes the Gr-systems such as Eqs. (5 and 6) which however in general does not allow the superposition of different waves. In this sense, calling them "nonlinear waves" may be justified; yet, we should clarify that each wave's dynamics is linear on its own.

For $\hat{u}_k \in \mathbb{C}$, the algebraically closed field of complex numbers, the existence of solutions to the complex algebraic equations implied by Eq. (8) and the corresponding Gr-system, is generally ensured by the consistency of the original system, and, as said earlier, analytical and numerical experiments show that such closed travelling-wave Gr-systems (second-order complex multivariate algebraic equations) admit solutions whose number increases rapidly with $K$. It would be nice to have a complete knowledge of the solutions' algebraic structure, or the geometrical structure of the algebraic varieties (see Ref. [22] and the comprehensive references therein, including the many original articles and monographs on the algebro-geometric approach, for constructing in physical space the KdV special solutions, including the travelling waves — more in Sec. II C), so that we can identify some system objects, like the (super)multiplets in particle physics. However, achieving this for general $K$ seems unattainable as of now. Note that the arbitrary phase or the freedom of $e^{\hat{i}k\tau_0}$ in $\hat{U}_k$ of the travelling wave mentioned in the above is not time-dependent, so we don't observe the nontrivial $U(1)$ group symmetry. However, intuition and numerical experiments suggest that truncation may introduce certain symmetries or structures within the Gr-systems. So, our strategy is to obtain insights based on example analytical solutions and numerical experiments analyzed with additional theoretical tools.

Indeed, numerical results show multi-longons and suggest a Kolmogorov-Arnold-Moser (KAM) mechanism at work, which should be underlied with presumably whiskered tori of multi-frequency orbits, in particular the quasi-periodic solutions. We then look for solutions of more general dispersion $\omega(k)$:

$$\partial_t \hat{u}_k = \hat{i}\omega(k)\hat{u}_k. \tag{9}$$

For fields composed of modes of different $|k|$s, the velocity $\omega(k)/k$ may vary with $k$ and can even be rationally independent, thus quasi-periodic. We defer more elaborated discussions to Sec. III C 4, after getting sufficiently familiar with more basic results.

Note that due to the (time-ordered) truncation, solving the travelling wave equation in physical space is not feasible, and writing down the dynamical system — namely, the ordinary equation of the waves for deeper analysis — is equally impossible, unlike traditional wave systems [15, 16]. Regarding the former challenge in physical space, where 'unusual' operators like the truncation $P_K$ obstruct conventional solutions (e.g., of travelling waves), we resort to work in the Fourier space to discover the waves; and, for the latter, appropriate approximation models are introduced.

The theoretical formulation and analysis in general assume $2\pi$-periodicity, while some numerical computations are performed with period 2 for different purposes. They are equivalent with re-scaling of the variables.

### A. The two branches of GrCP

#### 1. From compacton and peakon to longon

Exerting the pseudo-differential operator $P_K$ on the CP equation (1), the GrCP $\hat{u}_k$ fulfills

$$\partial_t \hat{u}_k = \sum_{p+q=k\in {}^K\mathbb{G}}^{p,q\in {}^K\mathbb{G}} \frac{-3\hat{u}_p q\hat{u}_q \mp 2p\hat{u}_p q^2\hat{u}_q \mp \hat{u}_p q^3\hat{u}_q}{\hat{i}(1\mp k^2)}. \tag{10}$$

This corresponds to augmenting the right hand side of Eq. (1) with ${}^K_{cp}g$, analogous to the $\alpha$- and $\beta$-terms in Eq. (2).

The combination of Eqs. (8) and (10) defines the GrCP travelling-wave ${}^K u^\#$ (below, the prescript '${}^K$' is neglected in general for simplicity).

For $\lambda = 0$, immediate examples include those occupying a single $|k|$ in $(K/2, K]$; while, for $\hat{u}^\#_k$ of two-mode occupation at $|k| = K/2$ and $K$, straightforward calculation leads to

$$3\hat{u}^\#_{\frac{-K}{2}}\hat{u}^\#_K = \lambda\hat{u}^\#_{\frac{K}{2}}, \tag{11}$$

$$\frac{4\mp K^2}{8}\hat{u}^\#_{\frac{K}{2}}\hat{u}^\#_{\frac{K}{2}} = \lambda(1\mp K^2)\hat{u}^\#_K. \tag{12}$$

Assuming $\hat{u}^\#_{\frac{K}{2}} = Re^{\hat{i}\Theta}$ with $R \geq 0$ and and the phase $\Theta$ taken to be 0 in the end for simplicity but without loss of generality, we have $u^\# = \frac{4|\lambda|}{3}\sqrt{\frac{2(1\mp K^2)}{(4\mp K^2)}}\cos\frac{K\zeta}{2} + \frac{2\lambda}{3}\cos K\zeta$ for $\mathrm{mod}(K, 2) = 0$, indicating no nontrivial stationary ($\lambda = 0$) solution. Note that, with $K = 4$ for instance, $\hat{u}_k$ of $|k| = 2$ and 4 do not interact to excite $\hat{u}_{\pm 5}$; that is, we can replace $K$ by $S = 4$, the occupied maximal $|k|$, and have actually $K = 5$, or, more generally for $u^\#$ occupying $L$ modes equally spaced in $|k|$, another parameter $S$ can be distinguished from $K$, with $\mathrm{mod}(S, L) = 0$, $S \leq K < (L+1)S/L$. When $L = S$, we may say that all $S = K$ modes are occupied; otherwise, there are unoccupied or empty modes. Particularly, here for $L = 2$, $S\zeta = \theta$ and $\lambda > 0$,

$$\frac{u^\#}{\lambda} = \frac{4}{3}\sqrt{\frac{2(1\mp S^2)}{(4\mp S^2)}}\cos\frac{\theta}{2} + \frac{2}{3}\cos\theta. \tag{13}$$

The $4 - S^2 = 0$ case for the compacton branch corresonds to trivial solution and is not essential. Actually, the corresponding issue already appears in Eq. (1) when the first two terms in the left hand side cancel each other for the mode of $|k| = 1$, which does not happen when all $|k|$ are rescaled to be greater than 1. For example, Fig. 1 presents the $u$-contours/carpets of the problem with the period $2\pi$ normalized to be 2, from the direct numerical simulations starting from the respective $u^\#$s at $t = 0$ with $\lambda = 1$, and the unnormalized $S = K = 4$ (for $2\pi$-period).

Both branches of GrCP ('GrCompacton' and 'GrPeakon') show accurate traveling-wave behaviors characterized by the respective $u^\#$ in the earlier stage, and instability and numerical errors (but with the dynamical invariants $\mathcal{M}_{CP}$, $\mathcal{E}_{CP}$, and $\mathcal{H}_{CP}$ conserved accurately — see the numerical method below) lead to states with interacting solitonic structures of clear periodic nature (akin to the well-known KdV quasi-periodic and almost periodic solutions), referred to as 'longons' (explained later), in a more stable state. As an effect of the truncation, which removes the singular behavior that would otherwise happen on the real line, both branches display similar qualitative and quantitative behaviors in such appropriately designed setups. Henceforth, we focus solely on the peakon branch, i.e., the GrCH.

Compacton and peakon correspond to different degrees of branch singularities [16], but the Galerkin regularization push the singularities into the complex plane where they are presumably closer according to the exchange between them by the transform mentioned in the introductory discussion. This, in a way, accounts for the proximity of the two GrCP branches observed above. Nevertheless, it remains unclear what factors determine the ultimate, more stable longon state. Could there be an unknown invariant at play? As we shall see, the GrCP scenario of the transition from the travelling wave to interacting longons is actually quite universal [17], suggesting probably some other mechanism.

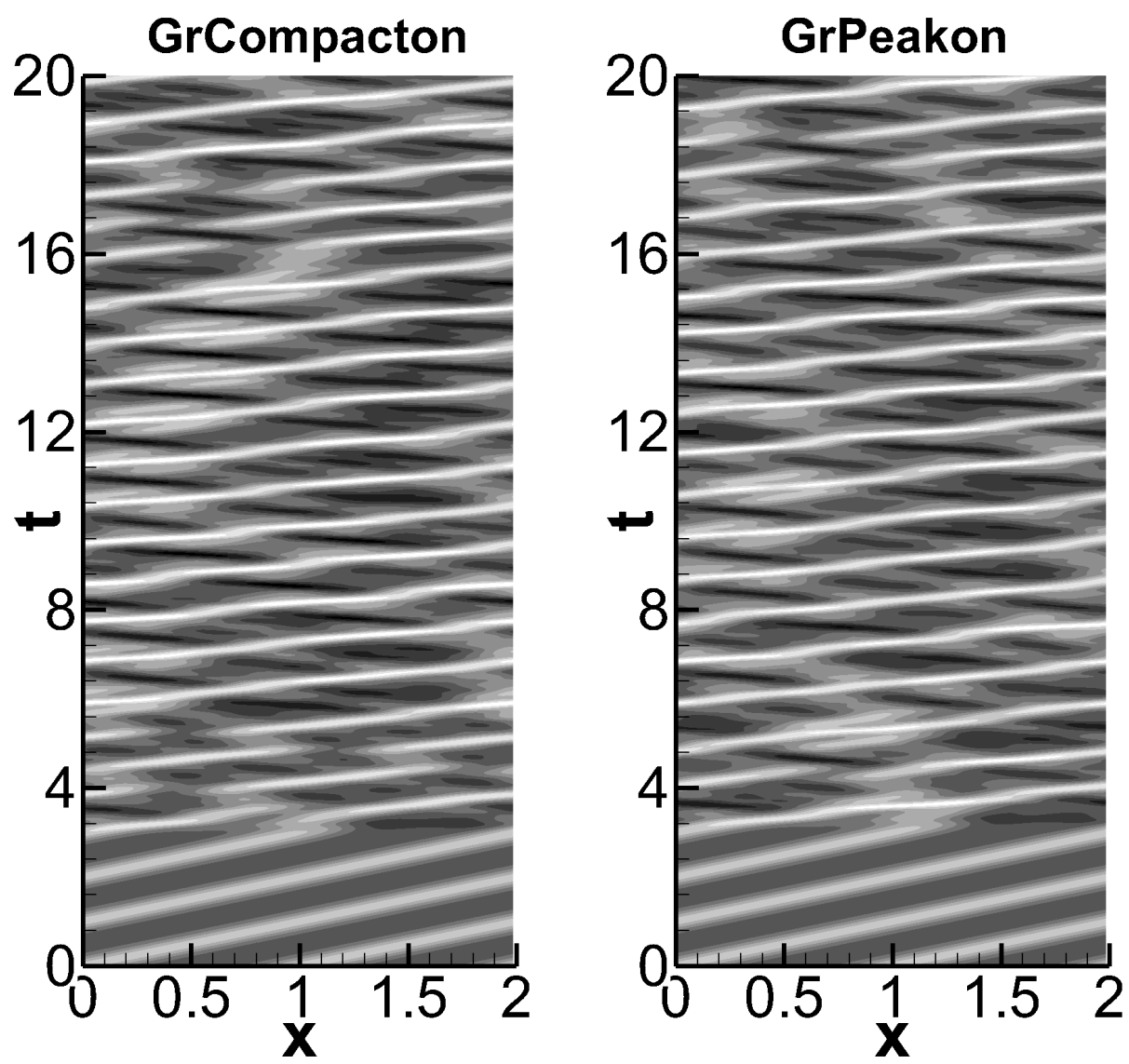

Figure 1. Comparison of the GrPeakon and GrCompacton contours/carpets: brighter color for larger values of $u$.

### 2. A linearly dispersive approximation of the truncation

We argue that the Galerkin truncation can be approximated using suitable linearly dispersive hKdV models. Specifically, this approximation should converge to the ${}^{K}_{cp}g$ (or simply ${}^{K}g$ when there is no risk of ambiguity). Whether and how the convergence towards the CP model occurs as $K \to \infty$ warrants a separate investigation. Additional remarks relating to the studies by Li and Olver [15, 16] will be provided later.

For an appropriate sequence $\omega_O$ of the dispersive functions $\omega(n)$ in the model $\hat{a}_n = -\hat{\imath}\omega(n)\hat{v}_n$ in Eq. (3), with $\omega_O(m) \to \infty$ for all $m \notin {}^{K}\mathbb{G}$ and $\omega_O(k) \to 0$ for all $k \in {}^{K}\mathbb{G}$, the corresponding hKdV model can be used to approximate the decoupled GrBH (sub-)dynamics with well-prepared $u_0$ in ${}^{K}\mathbb{G}$, and similarly for other Gr-systems. The asymptotic Gr-dynamics is supported by the fact that the intra- and extra-Galerkin frequencies can not match to form a resonant triad with a large jump of $\omega(n)$ in the classical resonant wave theory (e.g., Ref. [31] and references therein): the extra-Galerkin modes, if set up initially ('ill-prepared'), however, can have their own dynamics, not of the interest here though. For understanding some physics of dissipation, a choice of $a$ in Eq. (3) in Ref. [32] was the dissipation function $\propto -(k/k_G)^{2O}$ ($K < k_G < K+1$) for integer $O \to \infty$, but more consistent with the current situation is the dispersive model such as the hKdV

$$\omega_O = \begin{cases} (\frac{k}{k_G})^{2O+1} & \forall\ k \notin {}^{K}\mathbb{G} \\ 0 & \forall\ k \in {}^{K}\mathbb{G} \end{cases} \quad (14)$$

The even-order hyper-dispersive Benjamin-Ono type models, among others with appropriate dispersive speudo-differential operators, also works, but it makes no sense to go through various models for our purpose here. It is intuitively clear and also possible to be made rigorous in nonstandard analysis [26] that ${}^{K}\hat{g}_m/\hat{u}_m$ for infinitesimal $\hat{u}_m$ with $m \notin {}^{K}\mathbb{G}$ is in general a (hyper)complex number, which is why both the dissipative (in Ref. [32]) and current dispersive models, corresponding respectively to different choices of appropriate infinitesimal $\hat{u}_m$s, converge to the truncation; however, establishing the rigorous results in a unified way systematically with nonstandard analysis is beyond the scope of this work, deserving a separate special study. The (linearly) dispersive approximation does not have the difficulty of nonconservation, or the necessity of additional forcing for balance, as the dissipative one for any finite $O$, which is crucial for the following discussions.

The different but consistent perspective here, particularly associated to the invariant (whiskered) tori corresponding to the longulent state, is the persistence or KAM theorem which will be futher remarked on (Sec. III C 2).

Note that for our model, the approach of Ref. [15] would lead to an $(O+1)$th-order or $O$-variable dynamical system. Reverting Eq. (10) back to the CP equation by removing the brute-force truncation but then adding the above linear dispersion, we have

$$\partial_t \hat{v}_k = \sum_{p+q=k} \frac{3\hat{v}_p q\hat{v}_q \mp 2p\hat{v}_p q^2\hat{v}_q \mp \hat{v}_p q^3\hat{v}_q}{\hat{\imath}(1 \mp k^2)} + \hat{\imath}(\frac{k}{k_G})^{2O+1}\hat{v}_k \quad (15)$$

or

$$v_t \pm v_{txx} + 3vv_x = \mp 2v_x v_{xx} \mp vv_{xxx} + \frac{(-1)^O \partial_x^{2O+1} v}{k_G^{2O+1}}. \quad (16)$$

For travelling-wave solution $v = \phi(\zeta)$ parameterized by $k_G$ the idea of dynamical-system analysis and extension to the complex plane [15, 16] can be carried over to the current

$$\lambda\phi' \mp (\lambda + \phi)\phi''' - 3\phi\phi' \mp 2\phi'\phi'' - \frac{(-1)^O \phi^{(2O+1)}}{k_G^{2O+1}} = 0. \quad (17)$$

Rather than the convergence to the compacton-peakon model with $k_G \to \infty$ for some fixed large $O$ (actually 'non-convergence' is expected, in contrast to Li and Olver's case [16]), we claim the convergence with $O \to \infty$ for some fixed $k_G$ to the GrCP system whose singularities are already in the complex plane. [Now, we can transform the question of large-$K$ limit of Gr-systems to the large-$k_G$ limit of the model (14), with $O \to \infty$ taken first, applied to the original systems. Note that the limits $\lim_{O\to\infty}$ and $\lim_{k_G\to\infty}$ may not be exchangeable. We will have more discussions of the large-$K$ limit for GrBH later.]

For clarity, we take $O = 2$ (quite small but sufficiently clear) for demonstration. Let us adopt Li and Olver's [15] convention, using Eq. (2) with $\alpha = \beta = 0$ and $\gamma = \pm 1$. Following their analysis, straightforward calculation tells us the fix point $(a, 0, 0, 0, 0)$ and the eigenvalues of the Jacobian matrix, telling that we have, near the fixed point, the center manifold, stable manifold and unstable manifold, and, presumably the homoclinic orbit (for the infinite domain problem) as the limit of the periodic solutions. The problem can be more conveniently discussed in the hKdV approximation of the simpler GrBH later, because it is essentially the same as the KdV case in Ref. [15].

### B. The general scenario

The GrBH $u^\#$ corresponding to the GrCP solution (13) for $\lambda > 0$ and $L = 2$ is

$$u^\# = 2\sqrt{2}\lambda\cos(\theta/2) + 2\lambda\cos\theta. \qquad (18)$$

Similarly, for $L = 3$, we can obtain a three-mode-occupation

$$u^\# = \lambda[-2\chi_1\cos(\theta/3) + \chi_2\cos(2\theta/3) - \chi_1\chi_2\cos\theta] \qquad (19)$$

where $\chi_1 = \sqrt{\frac{5-\sqrt{5}}{5}}$ and $\chi_2 = \sqrt{5} - 1$. Accordingly, ${}^K g^\# \propto S\lambda^2$; for example, corresponding to Eq. (18),

$${}^K g^\# = -(S\lambda^2)[3\sqrt{2}\sin(3\theta/2) + 4\sin(2\theta)]/2. \qquad (20)$$

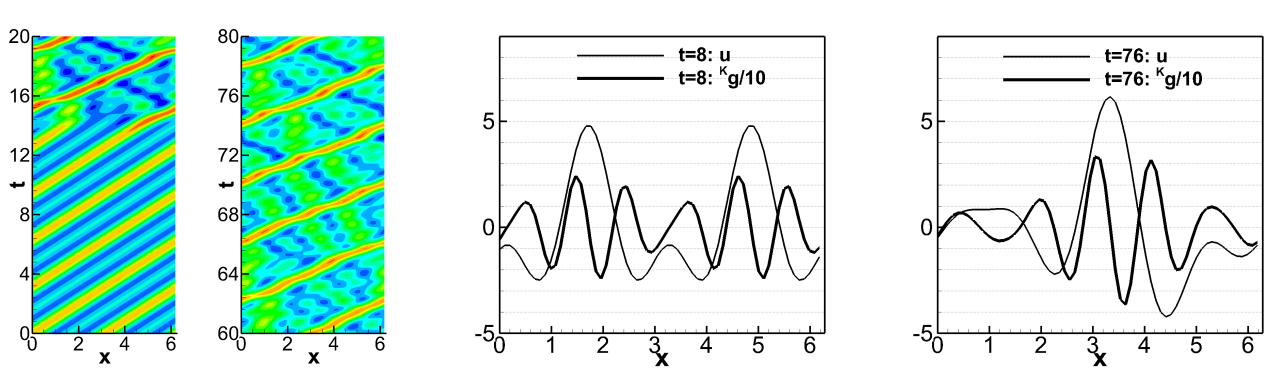

(a) Developing longons from solitary waves (left) and of developed ones (right).
(b) Snapshots of $u$ and ${}^K g$ at $t = 8$ of the accurately reproduced analytical solitary wave.
(c) Snapshots of $u$ and ${}^K g$ at $t = 80$ of the solitonic longon among chaotized ones.

Figure 2. GrBH numerical results starting from the solitary wave (18) with $S = K = 4$ and $\lambda = 1$: the color coding for the $u$-contours in (a) can be read from (b) and (c).

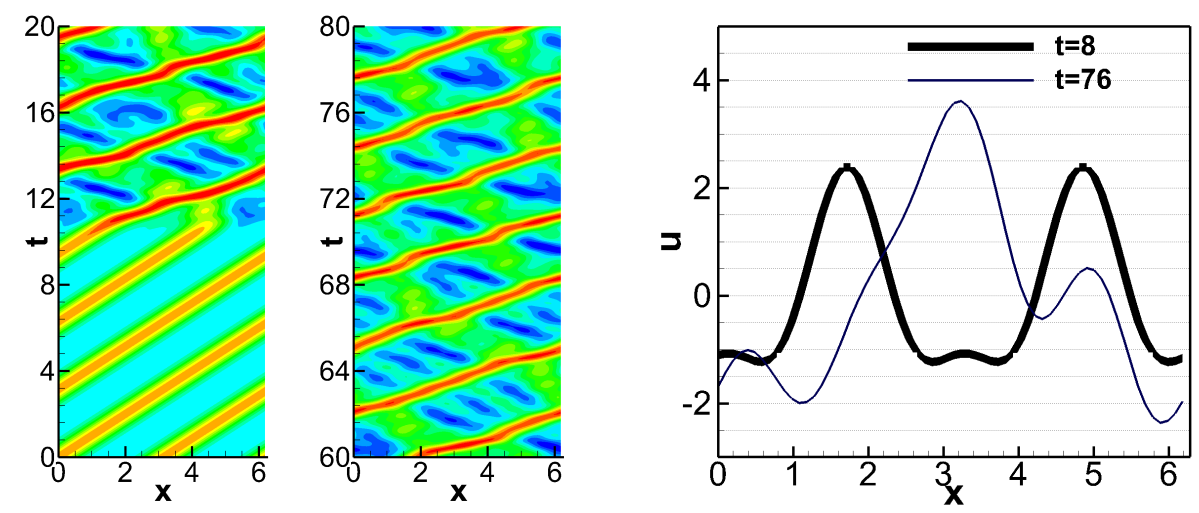

(a) $u$-contours of developing new longons from solitary waves (left) and of well-developed ones (right).
(b) Snapshots of $u$ and ${}^K g$ at $t = 8$ of the accurately reproduced analytical solitary wave.

Figure 3. GrCH numerical results starting from the solitary wave (13) with $S = K = 4$ and $\lambda = 1$.

Fig. 2 show some numerical results starting ($t = 0$) from the solitary wave (18), and, for better understanding from more cross-comparisons, the peakon branch of GrCP, i.e., GrCH results are presented in Fig. 3.

Large-$L$ travelling-wave $u^\#$s may be obtained, with the help of numerical approximations in various approaches (such as Monte Carlo, symbolic computation, and also the functional minimization used by Hyman in [21] for other special KdV solutions); e.g., for $L = 4$, a set of approximate $\eta$s in

$$\frac{u^\#}{2\lambda} \approx \eta_1\cos\frac{\theta}{4} + \eta_2\cos\frac{\theta}{2} + \eta_3\cos\frac{3\theta}{4} + \eta_4\cos\theta \qquad (21)$$

were found. Specifically, $\eta \approx \{-0.507, 0.450, -0.376, 0.292\}$ in the above ansatz, among others such as the seven-active-mode solution with $\eta \approx \{-0.288, 0.272, -0.250, 0.225, -0.196, 0.165, -0.056\}$, correspond to GrBH solitary waves resembling the cnoidal wave familiar in the periodic KdV and Toda lattice models, except for weaker wiggles between the strong pulses, which is also the case for Eqs. (13, 18 and 19). The wiggle counts in such waves increase with the mode numbers.

Increasingly diverse travelling-wave shapes, combinations of various pulses, emerge for higher $K$ values. However, numerical experiments seem to indicate that, while indeed the transient states appear to have (roughly) such shapes of longons, few of them (actually so far only one in the precise form of them — the Mexican-hat, i.e., a strong pulse along side with weaker wiggles) are clearly seen in the developed stage as the major longon profile. As an illustration, Fig. 4 shows a travelling wave occupying four modes, with the specific $\eta$ given in the above, and its ultimate outcome: although the developed longons do somehow inherite some features of the corresponding $u^\#$, their strongest pulses lack key characteristics, such as the 'camel-back' shape, compared to the original.

For other allowable $K$ values [$< (L+1)S/L$] for a given $S$, while the solitary wave and the big scenario are the same, the later-time developments of instabilities and longons are quantitatively different: Fig. 4 shows that a larger $K$ actually leads to more stable evolution and different developing and developed longon profiles.

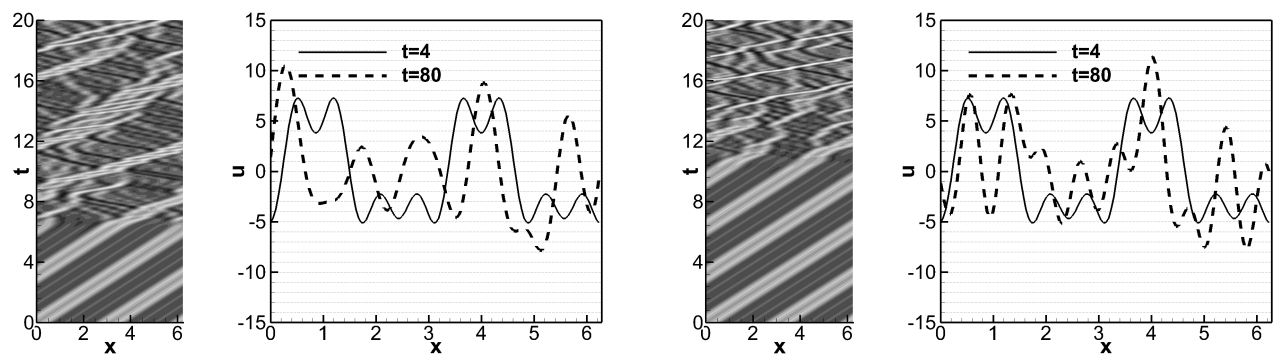

(a) $K = 8 = S$. (b) $K = 9 = S + 1$.

Figure 4. Evolution of the GrBH fields corresponding to Eq. (21).

For the hKdV model (14) added to the BH equation, we have the dynamical equation for the travelling-wave solution,

$$-\lambda\phi' + \phi\phi' + \phi^{(2O+1)}/k_G^{2O+1} = 0. \quad (22)$$

Proceeding with the standard dynamical system approach for KdV [15], extended to higher orders/degrees, we have the stable manifold, unstable manifold and center manifold, along with the homoclinic orbit, for the infinite-domain problem as the limit of periodic solutions. [The differences between even and odd $O$ in the details diminish as $O \to \infty$ with the absolute eigenvalues approaching $k_G$. We may choose the sub-sequences of, say, odd-$O$ approximations for cleaner or simpler analysis, if needed.] In this sense, it should be appropriate, even by more rigorous mathematical standards, to claim that solitary-wave solutions exist in GrBH.

### C. Universality and beyond

High-dimensional Gr-systems, such as the truncated Kadomtsev-Petviashvili and Zakharov-Kuznetsov models, and their nonlinearly dispersive versions [11, 33], similarly admit solitary-wave solutions and, presumably, the longons from the interaction potentials, ${}^K\mathcal{G}$. Actually, we have also found longulent states in the GrNLS [34]. It is not impossible that new surprises can emerge from more complex multi-component Gr-(magneto)hydrodynamics which Lee [5] initiated with the assumption of generic thermalization but did not explore the possibility of the travelling waves, quasi-periodic solutions and their implications (see the extension of the CH nonlinear dispersion to modify hydrodynamics [35] and the most recent KAM proof of the quasi-periodic vortex-patch solution to the Euler equation [36]).

A complete understanding of the mechanism of the universal presence of longons among the models we have tested (including GrKdV and GrRH [17]) is the furhter work underway, including the (a-posteriori KAM proof of the whiskered tori, the rigorous details definitely belong to another specialized communication. One possibility is the existence of appropriate additional invariants, which constrains the mode interactions and geometrical configurations. The rugged dynamical invariance of $\mathcal{H}$, the truncated version of $\mathcal{H}_{BH}$ for GrBH, and similarly, the truncated $\mathcal{H}_{CP}$ for GrCP are believed to play important roles in the formation of longons (more in Sec. IV). We should turn to more specific results to accumulate solid information for insights: later in Sec. III C 4, we will come back to more theoretical considerations with the idea of on-torus or torus-specific invariants for quasi-periodic tori associated to the KAM arguments.

We will henceforth focus on the minimal GrBH case for further expedition.

Note that we have not studied the multiple-mode stationary $u^\#$ with $\lambda = 0$. One of the reason is that it appears natural to introduce them to address the relevant issue associated to the 'Hamiltonian effect' later.

Constrained variation,

$$\delta(\mathcal{H} - \lambda\mathcal{E})/\delta u = 0 \Longrightarrow P_K(u^2)/2 = \lambda u, \quad (23)$$

thus also the $u^\#$ satisfying Eq. (8), i.e.,

$$\partial_t \hat{u}_k^\# = -\frac{\hat{i}k}{2} \sum_{p,q,p+q=k\in{}^K\mathbb{G}} \hat{u}_p^\# \hat{u}_q^\# = -\hat{i}\lambda k \hat{u}_k^\# \quad (24)$$

which leads us to considering the Hamiltonian effect to which the observation of the failure of strong mixing observed in some GrBH data were attributed [29]. But, we will show that zero-Hamiltonian data can also lead to solitonic structures. It turns out that (quasi-)piecewise-constant data also play a unique role in the nonlinearly dispersive effects of the truncation.

The term 'Hamiltonian effect' is more of physical flavor. Mathematically, the above constrained variation and the resulting travelling wave solutions are a particular type of special solutions. We have set $\mathcal{M} = 0$ which even if included in the above constrained variation does not affect the travelling-wave result. This is like the well-known KdV case where we have infinitely many invariants and where the corresponding variation extremizing the Hamiltonian, associated to $J_{KdV}$, constrained by lower-order invariants also results in the travelling wave, i.e., the cnoidal wave, a particular one among others from different constrained variations corresponding to higher-order KdV equations [20, 21]. Here no invariants of even higher orders are known to be rugged with Galerkin regularization (thus no "higher-order GrBH equations"), but still the angle of view in terms of special solutions defined by the invariants may be associated to the interacting longons.

The "universality" in Sec. II C is also in the sense that the major longon features from low $K$ (= 4 and 6 examined in the above so far) extend to large $K$s, which merit further investigation along with other aspects. Particularly, the fact that $u$ in the developed longon state, with disordered components though, is still separated (at least partly) may escape from our attention for small $K$s (Figs. 4, 3 and 2), but will become more apparent for larger $K$s: for instance, $u(t > 77)$ in Fig. 6 below never acquires the values around 20, the latter being between $\max u$ and $\min u$ though, thus is separated, at least partly in the interval, say, $(15, \max u)$.

## III. FURTHER EXPEDITION

As mentioned, further demonstration will be restricted to the GrBH case.

### A. Methods

None of the aftermentioned tools are required to be well mastered to reasonably understand the results here for general audiences (leaving the specifically involved aspects, such as the KAM proof, for elsewhere), although mathematical tools for theoretical guidance include the Hamiltonian method, KAM theorem, beside others such as the basic knowledge of the (weak-solution) theory of partial differential equations (particularly for the case of piecewise-constant data).

Numerically, the standard pseudo-spectral method is ideal for our periodic problems, and the classical fourth-order Runge-Kutta scheme generally provides adequate time accuracy with reasonable sized of time steps. However, for specific needs, additional techniques like exponential time differencing [37], for the hKdV model (14), and numerical symmetries, to confine solutions to invariant lower-dimensional sub-manifolds, are employed in certain computations as appropriate.

#### 1. The (pseudo-)spectral method

Let the periodic lattice coordinate satisfy $x_j = x_{j+N}$, whence $v(x_{j+N}) = v(x_j) =: v_j$ for $j = 0,\ 1,\ 2,\ ...,\ N-1$, defining a discrete torus $\mathbb{T}_N$. The theoretical foundation (also used in Ref. [24]) of the (pseudo-)spectral method and the lattice representation of the Gr-continuum lies in replacing $\hat{v}_k$ defined earlier by the discrete Fourier transform (DFT) for $|k| \leq M$ (with $N-1 = 2M$ here), $\hat{\hat{v}}_k := \sum_{x_j\in\mathbb{T}_N} \frac{v_j}{N} e^{-\hat{i}kx_j} = \hat{v}_k + \sum_{i\neq 0} \hat{v}_{k+iN}$. The aliasing error, represented by the second term, can be mitigated using dealiasing techniques like zero-padding or, alternatively speaking, truncation at $K < N/3$ ("2/3-rule"). Unifying the dealiasing and the Galerkin truncation results in, correspondingly, $\hat{\hat{u}}_k = \hat{u}_k$ for $u = P_K v$ in the GrBH equation (5), i.e., $\partial_t u_j = -P_K \partial_x u_j^2/2$, so

$$\partial_t u_j = \sum_{p+q=k\in{}^K\mathbb{G}}^{p,q\in{}^K\mathbb{G}} \sum_{x_m\in\mathbb{T}_N}^{x_n\in\mathbb{T}_N} \frac{k u_m u_n e^{\hat{i}(kx_j - px_m - qx_n)}}{2\hat{i}N^2} \tag{25}$$

where the right-hand side in physical-space variables reveals the GrBH lattice dynamics explicitly.

The 2/3-rule ensures sufficient sampling with $N$ sites for the Gr-continuum of $2K+1$ degrees of freedom, rendering extra sites (e.g., doubling $N$) dynamically redundant, unlike conventional lattice models [38]. The pseudo-spectral method marches in Fourier space, evaluating the nonlinear term in physical space via DFT of $P_K u_j^2$, $\partial_t \hat{u}_k = -\frac{\hat{i}k}{2}\sum_{j=0}^{N-1} \frac{P_K(u_j^2)}{N} e^{-\hat{i}kx_j}$. So, the computation aligns precisely with the Gr-system with only errors from the computer roundoff and time discretization.

#### 2. The initial data

The initial data can be given either for $v_0$ or directly for $u_0$. For (quasi-)piecewise-constant case, preparing $v_0$ is practically more convenient, while it is more straightforward to prepare $u_0$ if the Hamiltonian should be controlled. Also, preparing $u_0$ helps offering infomation in Fourier space. For example, in Fig. 5 (b), the quasi-piecewise-constant $u_0 = P_K v_0$ with

$$v_0 \sim \sum_k \frac{-2\hat{i}}{(2k+1)\pi} \exp\{\hat{i}(2k+1)(x-\pi/2)\} \tag{26}$$

tells us that the Fourier coefficients are purely imaginary up to a $\pi/2$ coordinate shift, which can be used for controlling the errors or chaoticity.

a. (Quasi-)piecewise-constant data For numerical convenience, we prepare approximation of the piecewise-constant $v_0$ with three collocation points for each discontinuity, and the central point acquires the arithmetic mean of the left and right ones for their respective pieces, which is consistent with the Fourier analysis for obtaining $u_0$ and with the Rankin-Hugoniot (anti)shock velocity $u_s$ for the quadratic nonlinearity in the problem.

b. Large- and vanishing-Hamiltonian data $\hat{u}_k^\#$ in Eq. (24) can in principle be obtained by writing out all $N_k$ triads satisfying $p+q=k$ for each $k$, which, in practice, however is not trivial for large $K$. Instead of the brute-force computation, we observe that $\hat{u}_k^\#$ being of a real value uniformly over $k$ would solve the problem if we had for all $k$ the same $N_k$ (with $\mathcal{M} \equiv 0 \equiv \hat{u}_0$, or $\hat{u}_0$ truncated), which can be realized by some different truncation with the Galerkin space other than the ${}^K\mathbb{G}$ that is concerned in this work; now $N_k = 2[K + \mathrm{sgn}(|\mathcal{M}|)] - 1 - |k|$ changes relatively slow with $k$ for reasonably large $K$ ($> 10$, say). So, in general, all Fourier coefficients being the same value does not extremize $\mathcal{H}$ to have travelling waves but results in ansatz reasonably close to $u^\#$, which is precisely what we are interested in (since we already know what essentially the scenario is for the exact travelling-wave initial data). Also, from Expression (7) for the Hamiltonian, we see that if all the Fourier coefficients of $u$ are real and of the same sign, they accumulate most effectively for $|\mathcal{H}|$; and, if all the Fourier coefficients of $u$ are the same pure imaginary number, we have $\mathcal{H} = 0$.

The above analysis leads to the consideration of two $u_0$s composed of, respectively, all-cosine and sine modes, both with the same weights for each mode as good exam-ples; that is, respectively,

$$u_0 = \sum_{k=1}^{K} \cos(kx)/\sqrt{K} \tag{27}$$

and

$$u_0 = \sum_{k=1}^{K} \sin(kx)/\sqrt{K}. \tag{28}$$

#### 3. The truncation wavenumber

The collocation point numbers are taken to be powers of 2, so setting the Galerkin truncation wavenumber $K = N/4$ would be most economic to compute ${}^{K}g$ whose wavenumber is no larger than $2K$. However, in general, $K$ is set to be $\lfloor N/6 \rfloor$, the integer part of $N/6$ (85 for $N = 512$ as the setup by default below, unless otherwise specified), to examine and compare with various numerical experiments, say, using the linearly-dispersive model with different parameterizations which in some cases need more than $(1/3 - 1/4)N$ modes beyond $k_G$. Actually, to have smoother outputs of the profiles in verifying some detailed properties (say, of ${}^{K}g$), some simulations with more severe truncations, such as $N = 2048$ for $K = 85$ were also used.

#### 4. The spectral quantification

Besides the conventional energy spectrum $E(|k|) := \langle |\hat{u}_k|^2 \rangle$, we may define

$$H(|k|) := \sum_{p} \langle \hat{u}_p \hat{u}_{k-p} \hat{u}_k^* + c.c. \rangle / 6 \tag{29}$$

with $\langle \bullet \rangle$ for time averaging. The energy transfer rate is $T(|k|) := \hat{i} \sum_p \langle \hat{u}_p \hat{u}_{k-p} \hat{u}_k^* - c.c. \rangle / 2$, showing some duality with $H(|k|)$. In GrBH absolute (statistical) equilibrium, $T = 0$ marks the balance of energy transfer, but $H$ provides additional insights into the structures.

Similarly, we may introduce the space-time or wavenumber-frequency spectra, as will be further remarked later.

### B. Invariant sub-manifolds and error control

The well-known invariant sub-manifold problem is indeed present in GrBH dynamics, as we have already shown analytical travelling-wave solutions for the sub-manifolds of various Gr-systems [see, e.g., Luce [39] for the dynamics of Galerkin regularized complex Ginzburg-Landau (GrCGL) system on sub-manifolds.] From the quadratic nonlinearity of the BH or GrBH equation, if the Fourier coefficients are initially purely imaginary or if the field is anti-symmetric about $x = \pi$ (with appropriate coordinate shifts), they remain so throughout the dynamics. Consequently, numerical errors arising from deviations from such symmetry can be efficiently eliminated in the (pseudo-)spectral method by setting the real parts to zero in Fourier space, which is unlike the GrCGL of very small degree of freedoms and with damping [39] (see also Ref. [34]). This technique is also applicable to cases with initial conditions such as $u_0 = \sin(x + \phi)$, particularly for $\cos x$ with a phase shift of $\phi = \pi/2$. Such a technique offers different perspectives about the numerical errors and the chaoticity of the dynamics, complementing the conventional method of simply comparing the results with different sizes of the time steps.

Eliminating certain numerical errors, specifically those arising from anti-symmetry freedoms via the aforementioned technique, or equivalently, enforcing specific exact dynamical symmetries, generally doesn't alter chaos entirely. However, in some particular situations, it does (controlling or suppressing chaos) and aids in discovering unstable precise solutions. This is pertinent to our previously mentioned GrBH travelling-wave and stationary solutions. For instance, Sec. III C 3 a illustrates the transition to a stationary state from a single longest mode using the aforementioned technique. Nevertheless, applying this method to other multi-mode cases, like those with initial data from Eqs. (26) and (28), doesn't yield such stationary solutions but rather ones resembling the uncontrolled counterparts, suggesting distinct basins of attraction within the invariant sub-manifold.

### C. Further results and analyses

It may be more engaging to view this section's analyses and findings as a standalone or parallel piece (except that some concepts and results are not repeated), rather than simply an extension of the preceding, more straightforward sections which seem to falsely indicate a crystal clear project vision from the beginning and to trivialize the developments of the ideas and discoveries.

We analyze mainly four types of numerical results here to show the nonlinearly dispersive properties of ${}^{K}g$ and the corresponding longons: they are from (quasi-)piecewise-constant initial data (Sec. III C 1), from particularly designed large- and zero-Hamiltonian initial data (Sec. A), from the linearly-dispersive model for comparison (Sec. III C 2) and from variations of $K$ for some preliminary insights of the large-$K$ behaviors (including those of the stationary solutions: Sec. III C 3).

#### 1. (Quasi-)piecewise-constant initial data

We will examine three types of quasi-piecewise-constant (QPC) $u_0$ leading to different scenarios involving solitization and thermalization of longons.

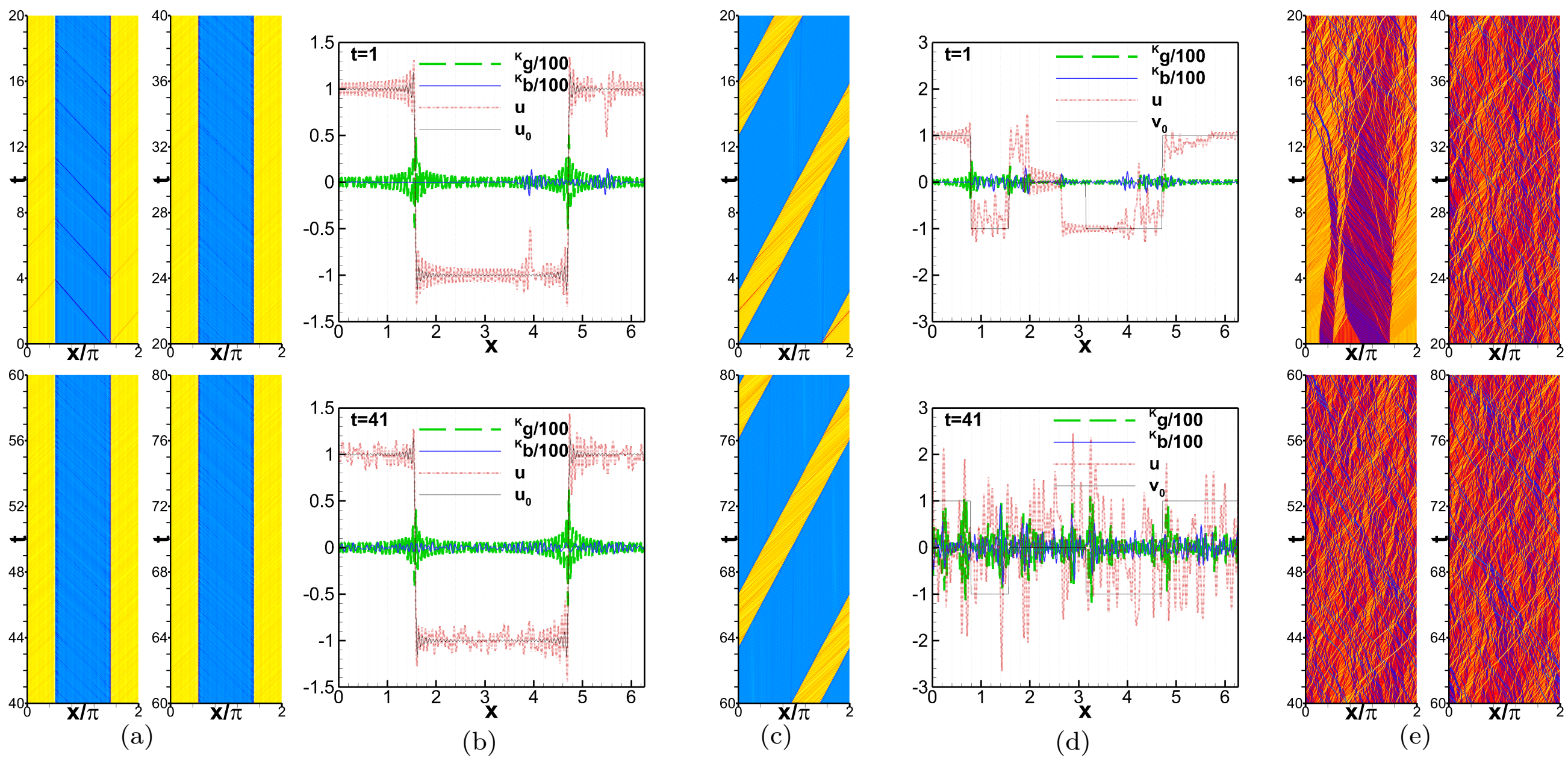

Figure 5. (a) The $u$-contours (carpets) in four space-time regimes for the case of a uniformly-zero shock velocity. (b) ${}^{K}b$, ${}^{K}g$ (rescaled) and $u$ snapshots at two times (respectively, upper and lower frames): the color-coding of (a) can be understood from the latter (with the common rule of brighter colors for larger values). (c) The early- and late-time $u$-contours for a case with a uniform but non-zero (anti)shock velocity. (d) ${}^{K}b$, ${}^{K}g$ and $u$-profile snapshots at two times (respectively, upper and lower frames) of a case with non-uniform shock velocities in $u_0$. (e) The $u$-contours from which the snapshots of (d) are taken.

a. Initial data with shock(s) and antishock(s) of uniform velocity: Eq. (26) explicitly expresses $v_0$ which is an exact stationary weak solution (but not the unique one) of the standard Burgers-Hopf Riemann problem for such a case. The speed $v_s$ of propagation of the discontinuity for appropriate weak solution is 0 for this case. With truncation and the Gibbs phenomena, $u_0$ is QPC with apparent oscillations, as we can see in Fig. 5 (b) and (d) where $u_0$ and $v_0$, respectively, are plotted deliberately for comparison, and the classical solution is unique. For a $K \to \infty$ limit with no oscillations in the initial data and the shock-antishock solution (not the unique one), the limit of the graphs (as Gibbs particularly distinguished from the graph of the limit [40] of the solution) presents overshoots and undershoots at the discontinuities located at $x = \pi/2$ and $3\pi/2$, but even more nontrivial is the $K \to \infty$ limit of $u$ from a finite-mode $u_0$ with oscillations, which is also related to the convergence problem of the (pseudo-)spectral method [23, 24] and will be further remarked later.

For the case of uniform nonvanishing (anti)shock velocity, snapshots are similar to those in (b), as indicated by the coded colors, thus not shown. The problem is trivially the same as that with vanishing (anti)shock velocity, by the Galilean invariance and the conservation of $\mathcal{M}$, but may be regarded as a more general case for discussing the results with those in the last paragraph together. The consequent space-time GrBH $u$-contours [carpets: Fig. 5 (a, c)] show strikingly clear persistence (selective nonthermalization) of the shocks and antishocks. Slight reduction of the (anti)shock strength in the beginning supports the excitations responsible for the "soliton-like" oscillations with clear space-time quasi-trajectories but, however, no clear recurrence or period. Such weaker structures [Fig. 5 (b)] eventually thermalize, now in the sense that they turn to appear random and statistically homogenized (the detailed characteristics may depend on specific initial data and will not be elaborated here), but still with clear space-time quasi-trajectories.

The persistent solitonic shock and antishock (a kind of 'soliton', loosely speaking), and, the soliton-like weaker oscillations are all called "longons", which will be eventually made clearer with more examples. (Some of) the weaker oscillations could actually also be solitons but just subject to so many collisions and thus the frequent and relatively severe phase shifts: a conceptual unification of the solitonic structure interactions and thermalization or chaotization is possible. Finally, note that ${}^{K}g$ is also strongest at the strongest shock and antishock longon locations, as in the solitary waves and in the interacting longons with much lower $K$s reported in previous sections.

b. Initial data with (anti)shocks of nonuniform velocities and of close strengths: Different (anti)shock velocities from $v_0$ lead to (anti)shock collisions as shown in Fig. 5 (d, e), and, the final "uniform thermalization" with no particularly persisitent but only soliton-like turbulent longons. The early structures are pre-thermalized but not solitonic longons. Large pulses of ${}^{K}g$ are also seen to stay with those strong longons.

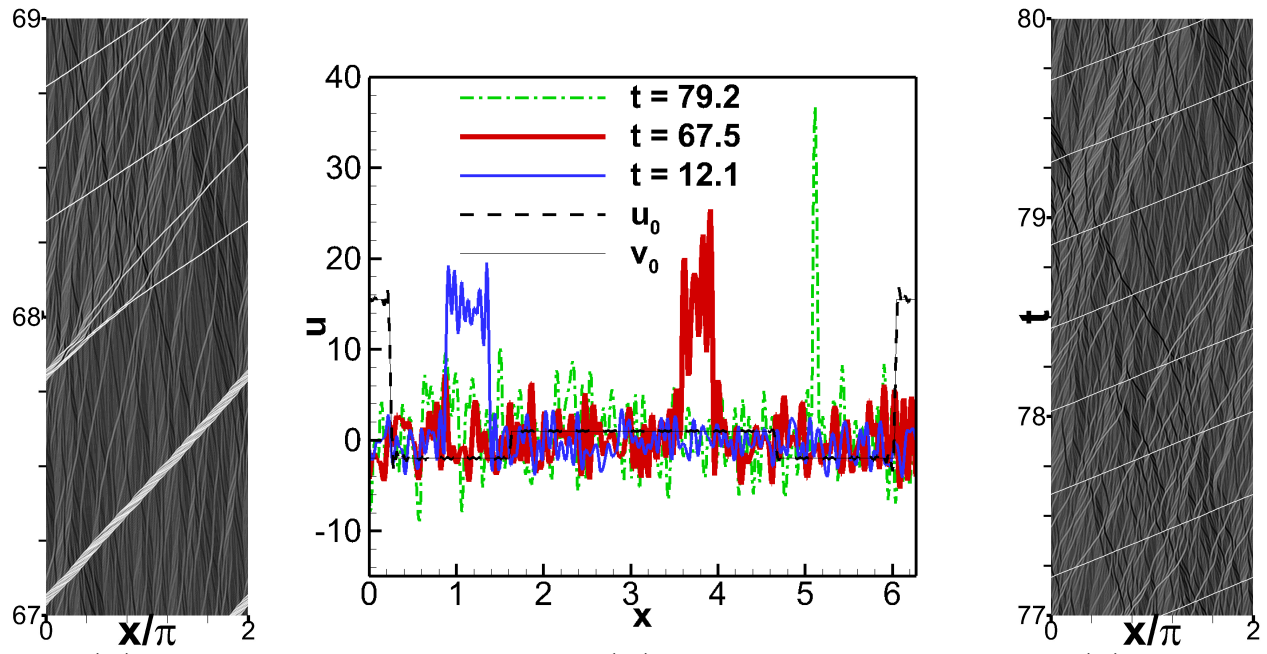

(a) (b) (c)

Figure 6. (a) Developing $u$-contours for the stage of solitization and thermalization with a late-blooming soliton. (b) $u$ snapshots indicating the whole process of solitization and thermalization. (c) $u$-contours for developed longulence whose color-coding, similarly that of (a), can be understood from (b).

c. Initial data with (anti)shocks of nonuniform velocities but of well-separated strengths: Fig. 6 presents the contours and snapshots of $u$-fields, including truncated $u_0$ and pre-truncated $v_0$ with the closest shock-antishock distance $\Delta x \approx 1.3$ and velocity difference $\Delta v_s \approx 16.0$ read directly from $v_0$, showing that the narrow piece has the shock and antishock sides persist for hundreds of units of the initial shock collision waiting time $\tau_{sc} = \Delta x / \Delta u_s \approx 0.08$: this piece deteriorates but well survives (c.f., snapshots at $t = 12.1$ and $t = 67.5$) up to about $t = 67.75$, leaving finally a much stronger single-peak solitonic longon (plotted for $t = 79.2$).

This case turns out to have nondimensional $h := \mathcal{H}^2/\mathcal{E}^3 \approx 1.99$ or, after Galilean-transformation to the coordinate with $\mathcal{M} = 0$, $\approx 1.49$ which would be a value large enough for the failure of strong mixing in Abramov and Majda's documemtation [29]. More meaningful notion of large Hamiltonian quantified by $h$ should of course be in the coordinate with $\mathcal{M} = 0$. Following the discussion of Sec. III C 1 a with zero-Hamiltonian data, we see that solitization or anti-thermalization does not need (large) Hamiltonian, but here and other various numerical experiments (not shown) with large-Hamiltonian QPC $u_0$ all similarly yield states with selectively unthermalized solitonic longon(s). We will come back to such an issue with on-torus or torus-specific invariants for quasi-periodic orbits, but it is appropriate to point out here that as indicated by the power spectra against $k$ (not shown here, but see the Appendix for similar results), the solitonic lognons are composed of dozens of wavenumbers of energy equipartition, with other dozens of wavenumbers of much lower energy; that is, the whiskered tori are of very high dimensions, dominated by dozens ($20^+$) of equipartitioned modes.

So, it appears that large Hamiltonian is sufficient but not necessary for solitization. That is, large Hamiltonian appear to assure structural properties of $u_0$ for the pulses and collisions appropriate for solitization, which is further demonstrated in the Appendix A for non-QPC data (together with other results, such as those of the hKdV model, the numerical accuracy and error control, and some details of the thermalized longons). This is similar to the results of periodic integrable systems, where, for instance, the "higher KdV equations" solved by the critical set defined through invariants correspond to the analogue of the (multi-)soliton state of KdV on the infinite line [20, 21].

### 2. Linearly dispersive approximation results

Also shown the Appendix A are the computations with Model (14) of Sec. II A 2, including also a case corresponding to zero-Hamiltonian data, with $k_G = 85.5 = K + 0.5$ and $O = 200$. To avoid the slow change for $|k|$ near $k_G^+$, $\omega_O(k)$ is emperically reset to be $-750\,\mathrm{sgn}(k)(|k| - k_G)$ if $(|k|/k_G)^{401} < 1300$, with the period normalized from $2\pi$ to 2.

The results indicate the convergence to GrBH dynamics with the Model (14) in the large-$O$ limit, as expected. Or, with the different (but consistent) perspective of this work, the persistence of the corresponding whiskered tori. The technical KAM proof at the level of methematical rigor belongs to another specific communication, but here we would like to point out that it is also possible to have a stochastic version of KAM theorem for the persistence of the longulent state associated to the whiskered tori even for the dissipative approximation, simultaneously with the dispersion of Model (14) or not, assisted with small random forcing for balance remarked in Sec. II A 2. That is, we may regard the dispersive and/or dissipative model, accompanied by the appropriate (random) forcing, as the perturbation to the clean Galerkin regularization/truncation, with some important structures preserved.

### 3. On the large-$K$ limit

The piecewise-constant shock-antishock in Sec. III C 1 a, Fig. 5 (a) and (c), in the corresponding $v_0$s are the BH weak solutions, which probably is responsible for the persistence of the QPC longons in the GrBH dynamics. Actually, for each $K$, we expect to find a stationary solution, with $\lambda = 0$ in Eq. (24), of the shock-antishock longon whose large-$K$ limti is that $v_0$, so that a more satisfying KAM understanding would be achieved. It is indeed possible to find such stationary $u^\#$ presenting the shock-antishock profile for small $K$ [17], but for general larger $K$ we do not have a theory for either ensuring the existence of shock-antishock stationary solution or identifying the (unique?) one from a large number of them. Actually, numerical efforts for this so far have neither provided any clear evidences.

Being not completely successful though, still, the efforts have been fruitfull. Here, some preliminary results from limited numerical experiments are offered for some large-$K$ insights into the persistence of the nonlinear dispersion and other more detailed properties of the longons at inifinitesimal scales.

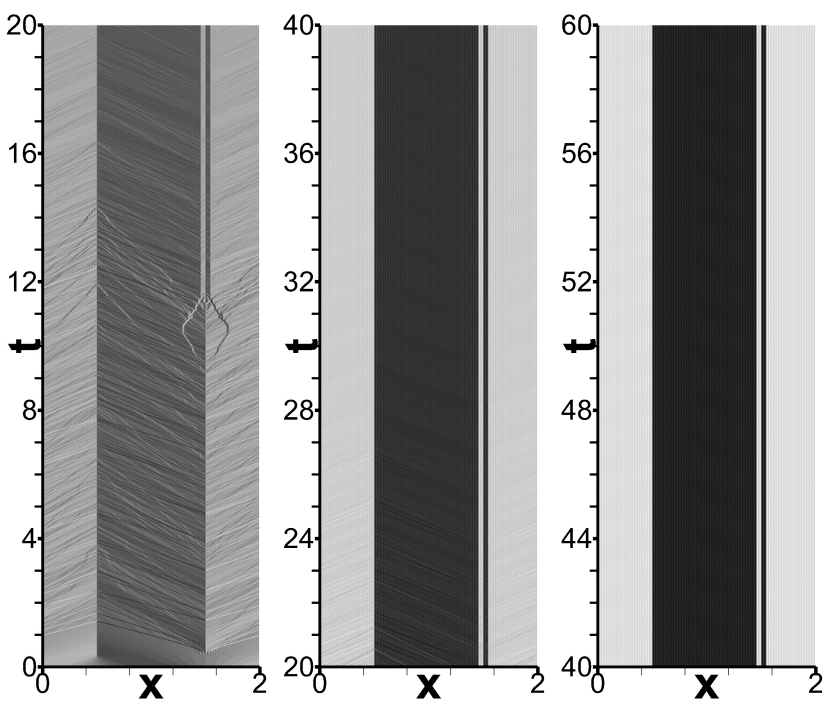

Figure 7. The space-time $u$-contours of the GrBH system with $K = 85$.

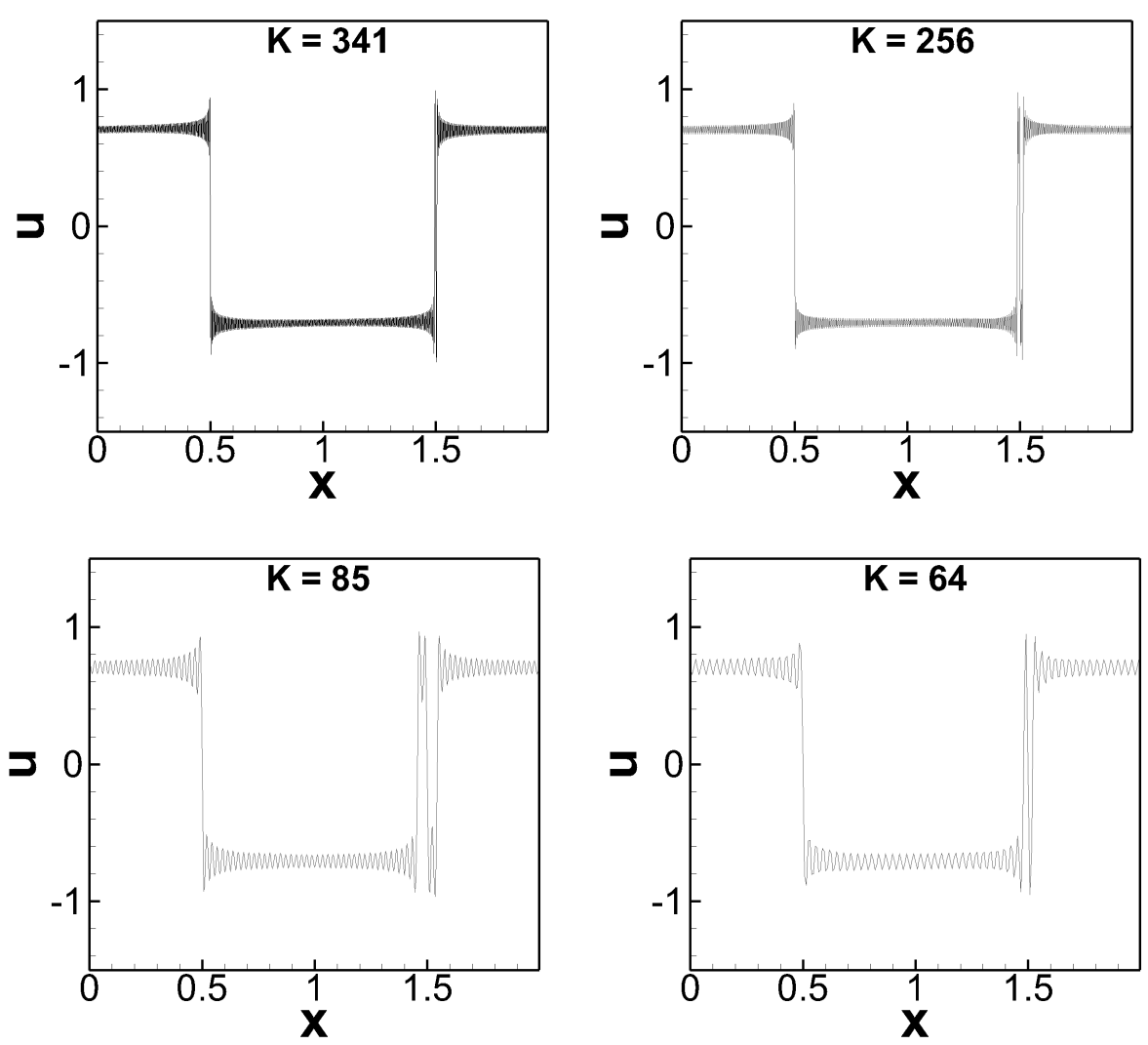

Figure 8. Stationary $u$-profiles with different $K$s from the same $u_0 = \cos(\pi x)$.

a. Stationary solutions: As said, $u_0$ truncated from the $v_0$ given by Eq. (26) allows error control with the analytical anti-symmetry condition imposed, which, unfortunately, does not lead to stationary but time-dependent solutions of the essentially the same feature as those in Fig. 5. Surprisingly, as shown in Figs. 7 and 8 for the $u$-fields from the computation of $x$-period 2, starting from the initial data $u_0 = \cos(\pi x)$ [phase-shifted from $\sin(\pi x)$ — Sec. III B] using the same error control technique for time up to $t = 60$, we see the evolution to the stationary solution. [Once the restriction on the invariant sub-manifold is relaxed, uniformly thermalized longons (not shown) similar to that of Fig. 5 (e) eventually present.]

The stationary $u$-profiles restricted to the invariant anti-symmetric sub-manifolds with larger and smaller $K$s (Fig. 8) from the same $u_0$ show a global large-scale picture similar to that in Fig. 5 (b), presenting however, instead of the anti-shock around $x = 1.5$, a big roll: the rolls have finer oscillations for $K = 85$ and 256 but not for $K = 64$ and 341 (checked by zooming in for the latter). Note also that the $K = 85$ has a wider roll than that of $K = 64$. For $K \to \infty$, sub-sequences of different details, as partly indicated by the variations of the profiles in Fig. 8, are possible, and there could be some $K$-dependent characterization of different sub-sequences. So, for the finiteness of $K$s, a conclusive inference about the evolution to a unique stationary solution or not can not be made here.

The local amplitudes of the stationary longons relative to the high and low pieces are decreasing but with increasing total variations (checked by observing that the amplitudes decrease in a way less fast than the increase of the periods). The finite data can neither say whether the local amplitudes of the oscillations vanish as $K \to \infty$, but if indeed, the total variations may still be “anything” (it is easy to construct function series of vanishing oscillation amplitudes but with nonvanishing or even infinite total variations.) If the oscillation amplitudes do vanish, we may envisage that there might be different sub-sequences with distinct $K \to \infty$ limits, sharing however the common interesting scenario with $b = g$, infinitesimal minor longons [26] around the major shock-and-roll longon and infinite total variations.

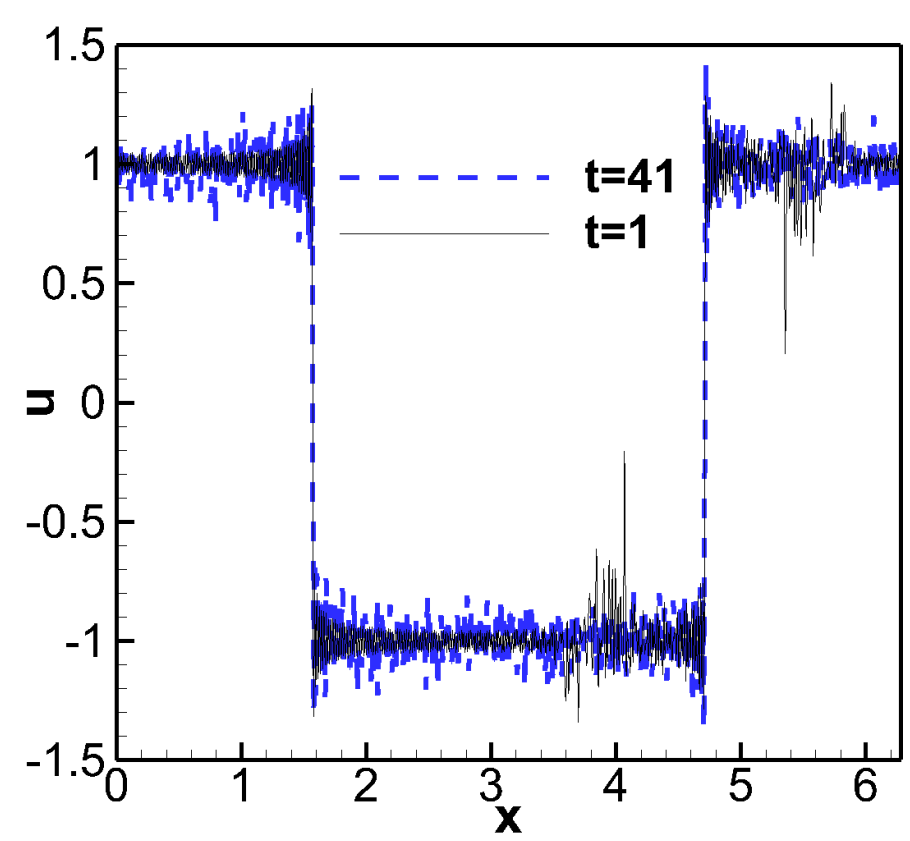

Figure 9. Quasi-piecewise-constant $u_0$ case for comparison with that in (b) of Fig. 5.

b. Nonstationary solutions: Fig. 9 presents the corresponding result from the same QPC $u_0 = -\hat{i}\sum_{k=-43}^{42}\exp\{\hat{i}(2k+1)(x-\pi/2)\}/[(2k+1)\pi]$ as that in Fig. 5 (b), but the GrBH is truncated at $K = 340$. As said, the error-reduction or chaos-controll technique used in last Sec. III C 3 a does not help here to reach a stationary state. The nonstationary minor longons away from the major shock-and-antishock longon have larger total variation (checked by examining the fine strutures) for such and other (not shown) larger $K$s, consistent with the estimation of Bardos and Tadmor [24] for the $K \to \infty$ limit.

## 4. Quasi-periodic solutions

We have seen the states with (multiple) solitonic longons and have mentioned quasi-periodic tori and KAM. The latter can be strengthened by considering the invariants and the solutions specified by a set of invariants, in a fashion similar to that of the integrable systems (such as KdV in, e.g., Refs. [20, 21]) with however the new notion of on-torus invariants. For instance, with

$$\mathcal{E}_\tau = \int_0^{2\pi} (\partial_x^\tau u)^2/(4\pi) dx \tag{30}$$

and $\mathcal{E}_0 = \mathcal{E}$, the critical set parameterized by the multipliers $\lambda_0$ and $\lambda_\tau$ of the variational principle

$$\delta(\mathcal{H} - \lambda_0\mathcal{E}_0 - \lambda_\tau\mathcal{E}_\tau)/\delta u = 0 \tag{31}$$

results in the replacement of the right-hand side of Eq. (24) by $-\hat{i}k(\lambda_0 + \lambda_\tau k^{2\tau})\hat{u}_k^\#$ and a solution, for $L = 2$ like (18), with initial phases $\theta_1$ and $\theta_2$,

$$u^\# = \sum_{z=1,2} Z_z e^{\hat{i}\frac{K}{z}\{x-[\lambda_0+\lambda_\tau(K/z)^{2\tau}]t\}} e^{\hat{i}\theta_z} + c.c. \tag{32}$$

which is quasi-periodic when $\lambda_0$ and $\lambda_\tau$, in $Z_1 = \sqrt{2[\lambda_0 + \lambda_\tau(K/2)^{2\tau}](\lambda_0 + \lambda_\tau K^{2\tau})}$ and $Z_2 = \lambda_0 + \lambda_\tau(K/2)^{2\tau}$, are rationally independent. Note that $\tau$ can be negative for $\mathcal{M} = 0$, with the understanding of Eq. (30) in terms of the pseudo-differential operator. For example, $\mathcal{E}_{-1} = \sum_{|k|=1,K} k^{-2}|\hat{u}_k|^2$. On such a 2-tori, according to Eq. (31) and the GrBH $u_t = -2\pi\partial_x(\delta\mathcal{H}/\delta u)$ with the Poisson bracket

$$\{\mathcal{F}, \mathcal{G}\} = \int_0^{2\pi} \frac{\delta\mathcal{F}}{\delta u}(-2\pi\partial_x)\frac{\delta\mathcal{G}}{\delta u} dx \tag{33}$$

inherited from KdV [27],

$$\frac{d(\lambda_0\mathcal{E}_0 + \lambda_\tau\mathcal{E}_\tau)}{dt} = \{\lambda_0\mathcal{E}_0 + \lambda_\tau\mathcal{E}_\tau, \mathcal{H}\} = 0. \tag{34}$$

The invariance of $\mathcal{E}_0 = \mathcal{E}$ means $\{\mathcal{E}_0, \mathcal{H}\} = 0$, so $\{\mathcal{E}_\tau, \mathcal{H}\}$ =0, i.e., $d\mathcal{E}_\tau/dt = 0$; also, from Eqs. (31) and (33),

$$\{\lambda_0\mathcal{E}_0, \lambda_\tau\mathcal{E}_\tau\} = \{\mathcal{H} - \lambda_\tau\mathcal{E}_\tau, \lambda_\tau\mathcal{E}_\tau\} = 0. \tag{35}$$

That is, we have the following

Theorem 1 (on-torus invariance) $\mathcal{E}_\tau$ ($\tau \neq 0$), $\mathcal{E}_0$ and $\mathcal{H}$ are mutually in involution on the torus associated to the GrBH critical set identified by $\delta(\mathcal{H} - \lambda_0\mathcal{E}_0 - \lambda_\tau\mathcal{E}_\tau)/\delta u = 0$.

Remark 1 In general, $\mathcal{E}_\tau$s in Eq. (30) do not mutually Poisson commute, neither for another choice

$$\mathcal{E}_\tau = \frac{1}{2\pi}\int_0^{2\pi} u^\tau/\tau! dx; \tag{36}$$

though, $\mathcal{E}_\tau$s from the latter with smooth (untruncated) $u$ do be invariant and mutually Poisson commute for the smooth untruncated Burgers dynamics.

A $\mathfrak{T}$-torus can be similarly realized with, say, $\mathfrak{T}$ different $\mathcal{E}_\tau$s or other on-torus effective invariants. $\mathcal{M}$ is not "effective" for GrBH, because the corresponding variation results in the multiplier, vanishing after after the spatial derivative when brought into $u_t = -2\pi\partial_x(\delta\mathcal{H}/\delta u)$. This is not the case for nonlinear Schrödinger (NLS) where the nonlinear term does not involve spatial derivative, and we can already have GrNLS 2-torus without introducing other on-torus invariant(s) [34].

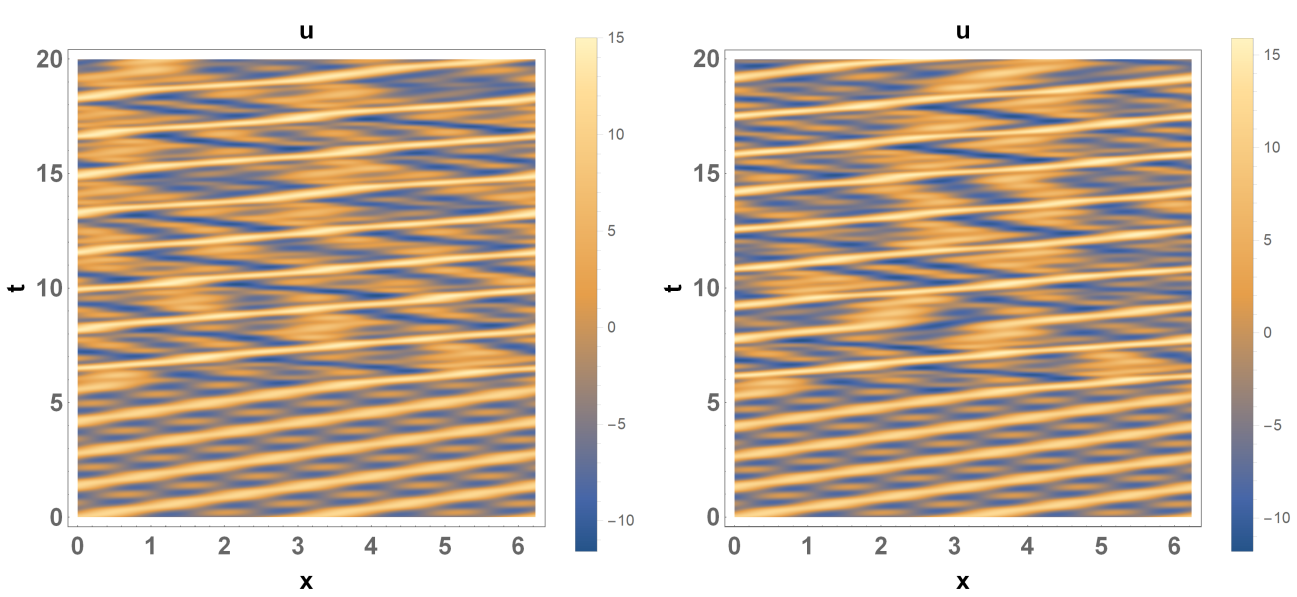

Figure 10. $u$ fields of two frequency vectors without (left) and with (right) the satisfaction of the Diophantine condition for the case of $K = 4$.

KAM analysis taking various conditions into account deserves a special study which is however beyond the scope of this note. The quite involved details cover issues such as the nondegeneracy condition (requiring the on-torus action-angle formulation) and the Diophantine condition. For example, we have tried a set of frequency vectors with, say, around $\lambda_0 = \sqrt{(6)}/10$ and $\lambda_1 = \pi/10$. Fig. 10 shows that the case satisfying the Diophantine condition (with $\lambda_0 = \pi/13$ and $\lambda_1 = \pi^2/30$) presents no particular better closeness of the initially quasi-periodic orbits than the periodic case (with $\lambda_0 = \sqrt{(6)}/10$ and $\lambda_1 = \pi/10$, rounded to one decimal place). This is the typical situation accordingly to our numerical experiments with various $K$s and the associated frequencies, rationally dependent or not, Diophantine-conditioned or not, which means, on the one hand, that both apparently show the breaking down (markedly at around $t = 5$, as seen by our bare eyes) of the exact solution and the emergence of (solitonic) longons, and on the other hand, that additional insights are needed to know precisely what (on-torus) invariants should be responsible for a specific longon state and how persistence would be working. Note that the late attractors may actually be "close" to the initial tori in the high-dimensional phase space, and that even Cantorized tori can still preserve continuity and closeness of the geometry to the initial ones: we can imagine that our initial tori is actually the projection of higher tori with zero components of frequencies which become nonzero but very small in the longulent state. It would be possible to obtain further insights for rigorous analysis by fermorming comparative exploration of various invariant tori with differrent on-torus invarants with phase-space analysis techniques, such as measuring the Hausdorff distance between structures etc.

We remark that the above quasi-priodic case can not be inifintely accurate due to computer accuracy, and might be worried that computer truncation would actually lead to periodicity. However, as shown, for instance, by recent numerical quantum revival and fractalization results [2, 9, 41], the properties of irrationally of the numbers can be well captured by normal numerical computations (say, with double precision).

The longulent state do appear closer to the exact two-frequency (quasi-)periodic solution than any others we obtained from the travelling-wave-solution data. Note that the numerical perturbations here originate from the computer roundoff, details of course depend on the perturbations, without however changing the overall scenario (as long as small enough). Although $\mathcal{H}$ and $\mathcal{E}$ are checked to be conserved with high precision as in all other simulations, $\mathcal{E}_1$ is not, being highly sensitive to the minute deviations from the precise orbit, thus not really the robust on-torus invariant responsible for the final state, and indicating the subtleness of the theory and numerics. But here the favorable physical picture naturally follows: the torus-specific invariants should be of different degrees of robustness (decreasing with $\tau$) which is associated to stability, resulting in longons of different degrees of (dis)order.

Since disorder beyond quasi-periodicity appears apparent in longulence, we tend to describe the latter with "pseudo-periodicity" which is at most quasi-periodic (the disorder could be a disguise of quasi-periodicity, if indeed). Nonintegrable conservative systems may all admit pseudo-periodic patterns, and the above mechanism with on-torus invariants generic. For example, we can also use pseudo-differential operators to design revised KdV model with even-odd alternative dispersions for particular mathematical and physical purposes, which indeed leads to pseudo-periodicity [41] as the Gr-systems here.

## IV. DISCUSSION AND OUTLOOK

In both L. Onsager's letter to C.-C. Lin [10] and T.-D. Lee's article [5], these two Nobel Laureates indepently made the same assumption of thermalization for the Galerkin regularized Euler equation. Their viewpoint appears to have been adopted in essentially all publications of Gr-systems truncated from different hydrodynamic-type equations (e.g., Refs. [28, 42–45]), with the sole exception of Abramov and Majda [29] who documented the failure of strong mixing with large-Hamiltonian data, to the best of my knowledge. It is somewhat surprising that the FPU discovery led by another heavyweight Nobel Laureate and colleagues, E. Fermi et al. [4], and the subsequent revolution, one of the greatest in the last century, in nonlinear physics were not sufficient to seriously shake such a belief.

Realizing that the truncation is actually an additional 'balancing' nonlinear dispersion ${}^{K}g$ is the key for all the discoveries made here. The intuition [6] was that, while the one-dimensional 'nice' balance could be favorable (for no sufficient 'escaping' space), the high-dimensional multi-component Gr-systems such as that considered by Lee [5] should have little chance to sustain a persistent coherence (for general $K$). Without the practically working concepts and numerical supports (such as the traveling-wave solution, solitonic structure, and linearly-dispersive approximation), the 'balance' intuition however would be otherwise too weak for the breakthrough in ideas; and, indeed, we have seen that even the solitonic longons are accompanied by the less-ordered or chaotic-looking component, not simply the old FPU recurrence and KdV soliton paradigm. As for general high-dimensional and/or multi-component Gr-systems, the concern can be dated back to Derrick [46], but, in view of the existence of solitonic magnons in the Landau-Lifshitz equation and tensorial gravitational solitons etc., and now, further strengthened by the first-hand results reported here, the intuition is updated to expect a good chance of realizing longons, not trivial though.

As mentioned in Sec. II C, the emergence of longons might be related to rugged invariants involving other quantities. Onsager and Lee's thermalization postulation on Gr-Euler, which has not been found to conserve any other global quantity than the quadriatic one(s), can be valid generically if rugged invariant involving additional quantity is necessary for longon formation [the caveat of additional invariant was actually mentioned by Lee [5] who later worked on soliton (stars!), by the way.] However, as we have raised for GrBH, on-torus Gr-Euler invariant(s) might also enter to change the game.

On the other hand, the states with solitonic longons motivate a phenomenology of "stochastic self-resonance": the 'noise' [47], conserving the corresponding rugged invariant(s) in the pseudo-spectral computation, develops with the instability to a stage of appropriate resonance with the main 'signal', the whole system behaving then in the KAM fashion 'near' some invariant tori determined by the initial data.

### A. Further discussion

It should be emphasized that, the less-ordered or chaotic-looking longons are notably "long" for Gr-systems of reasonably large cutoff wavenumber $K$, while the solitonic ones are infinitely long, in terms of space-time quasi-trajectories: even for the uniformly thermalized GrBH longulence in, e.g., Fig. 5 with energy equipar-tition (see also Ref. [45]), typical longon space-time (quasi-)trajectory lengths are still about dozens of times of the correlation length $\sim 2\pi/K$, estimated from $s = \pi/K$ and $2\pi/K$ being the first and second zero points of the spatial correlation

$$\langle u(x)u(x+s)\rangle = \frac{\mathcal{E}}{K}[\sin\left(K+\frac{1}{2}\right)s - \sin\left(\frac{s}{2}\right)]/\sin\left(\frac{s}{2}\right),$$

computed by the Wiener-Khinchin theorem from the equipartitioned energy spectrum.

The applicability of the linear-dispersion approximation for ${}^{K}g$ makes the above result and, more generally, the unification of thermalized oscillations and solitonic interacting structures conceptually natural, and hopefully quantitatively approachable for the latter. The result also aligns with the imagery of ‘Long’ (the oriental Dragon in Chinese Pinyin), the force carrier for various interactions between, say, human and nature, embodying the transformation between finitude and infinity, maintaining order amidst potential chaos, etc. Thus, the term ‘longon’ encapsulates these aspects with unity.

Echoing the opening remark of the introductory discussion, effective field theories in particle and condensed matter physics frequently involve truncations, akin to the wavenumber restriction here. Such physical theories are in general careful about the cutoff symbol function $p$ (used implicitly or analogously) and sometimes particularly use techniques, such as the renormalization group, to exclude the effects of cutoffs or to ensure insensitivity to the precise cutoff value. If the procedures are inappropriate or physical limitations intervene, our findings on classical systems hint at possible novel (quasi-)particles, whether fictitious or genuine. One possibility is that the effetive theories predict wrong particles beyond their effective regimes. This is not overblown or contrived from the extreme event of a hard cutoff, but rather natural inference supported by the results of the model (14) associated to other soft cutoffs and by the expectation of quantized counterparts.

## B. Outlook

Unlike the well-known consequence of compacton- and/or peakon-emergence from other classical nonlinear dispersions, another type of state with smooth structures, specifically the solitonic longons, is supported by ${}^{K}g$, which opens up the possibility of yet other nontrivial effects of nonlinear dispersions (just as as indicated by the quantum revival and fractalization in the nonlinear regime [2] of linear dispersion). We have seen ‘less is more’ from the Galerkin regularization, but still more are expected.

For complex systems, dissipation and dispersion mechanisms can coexist in both linear and nonlinear forms. The presence of nonlinear dissipation, probably accompanied by nonlinear dispersion, can be anticipated in variations of Gr-sysmtems with schemes other than the simple Fourier Galerkin truncation (e.g., Ref. [42] and references therein). Consequently, the corresponding dissipative longons may arise (for applications of nonlinear dissipation in physics systems, see, e.g., Ref. [48]). If the system is not dissipated so that the evolution can continue indefinitely under given rules, just as J. Conway’s ‘game of life’, recurrent or permanent coherent structures in suitable situations can emerge. So, some remarks on the extension and development of the ideas and techniques to other systems follow.

Conservativeness or even time-reversibility may be maintained with driving and damping mechanisms, as long as the latter are appropriately balanced. Beyond the discussion of universality in Sec. II C, the type of systems like the time-reversible She-Jackson [49] and Gallavotti-Cohen dynamical ensembles [50] warrant consideration. Such models modify the original damping to balance the driving in such a way that the systems become dynamically time reversible and conserve the targeted physical quantities. Together, the driving and modified damping can act as a nonlinear dispersion, potentially leading to new nonlinear dispersion effects. Further Galerkin regularization is thus analogous to the GrCP situation so that, in principle, similar analyses can be performed, with prospects of discovering new solitary waves, quasi-periodic solutions and interacting solitonic structures, among others.

Finally, it is observed that the fact that the GrBH $u$ extremizing the Hamiltonian constrained by the energy is a travelling-wave field seems dual to the cnoidal-wave case of KdV [20, 21]. If the existence of a variational principle or critical set is necessary for the existence of (solitonic) longons, a mathematical foundation akin to that of the integrable-system (finite-gap) theory is expected, which justifies our introduction of the quasi-periodic tori with on-torus invariant(s), a possible generic theoretical basis for nonintegrable conservative systems. Finite-gap theory has been used to construct (dense) KdV and NLS soliton gases with a special assumption of the band-gap distribution [55], but nonintegrable soliton gases, and particularly longulence here, need a more general theory as well. Indeed, as mentioned earlier, pseudo-periodic patterns with solitonic structures among much less ordered components present also in other systems of dispersions defined by the even-odd alternative pseudo-differential operator [41]. For clearer clues to a possible universal mathematical method and physical mechanism, we also investigate GrNLS in another note [34].

# V. ACKNOWLEDGMENTS

The communications with J. M. Hyman have been motivating. The author warmly acknowlege the early interactions with U. Frisch, A. Majda, W. Pauls and I. Timofeyev, and he also thanks H.-Y. Jin and Z.-G. Wang for the discussions on the physical effective field theories.

## Appendix A: "Hamiltonian effect"?

In this section, the following two subsections will present dual cases corresponding respectively to Eqs. (27) and (28), i.e., a case with large Hamiltonian, resulting in multiple solitonic longons, and a zero-Hamiltonian case with sharp pulses in $u_0$ but equal strengths of opposite signs, resulting in uniformly thermalized longons. It will also be demonstrated that typical major longon in all developed states, uniformly or selectively thermalized, present similar internal structures.

### 1. Large-Hamiltonian case

With the rational offered earlier, $u_0$ given by Eq. (27) with, again, $K = 85$ and $N = 512$, and, thus very large $h \approx 20.75$, was used for computation and results are presented in Fig. 11 which shows that different longons solitize or thermalize through the interactions: $u$-contours in (a) show astonishingly clear characteristics of two major counter propagating and interacting solitonic longons (of roughly but not very accurately linearly proportional amplitudes) the stronger one of which, as also shown in (b) and (c) for the profiles, is particularly highly concentrated but smooth, like a classical soliton [3], and with wiggles on the two sides of the major pulse.

Other, albeit much weaker, longons exhibit soliton-like characteristics, as evident from the long streaks of their (quasi-)trajectories, despite significant relative phase shifts. Note that the interaction of the two major solitonic longons is extremely robust numerically, with tiny differences upon varying the sizes of the time step changed from, say, $10^{-6}$ (here) to $10^{-7}$ (e.g., the corresponding results in Ref. [17]).

Fig. 11 (b) corresponds to the upper-left zoom-in panel of (a), showing that the amplitudes of the (major) solitons slightly oscillate [51]. This 'breathing' mechanism may resemble the emission and reabsorption of sound waves in Bose-Einstein condensates with harmonic potential (Refs. [52, 53] and therein).

The oscillations of ${}^K g$ imply the scenario that fluid particles, which otherwise would fly freely as indicated by the material derivative $Du/Dt$ in Eq. (5), bounce back and forth when in proximity, at a distance of $\sim 2\pi/K$, due to the Galerkin interaction of potential ${}^K\mathcal{G}$ with the exchange of longons carrying the Galerkin force ${}^K g$, and we can carefully check and will see more clearly below that ${}^K g$ is indeed half-wavelength ($\sim \pi/K$) of and centering at the peak location of the major soliton: at the center/peak of the major solitonic longons, both $B$ and ${}^K g$ spatially 'synchronize' with it and precisely vanish, and both steeply grow to maximum amplitudes of opposite signs near the half-height locations (HHLs) on both sides of the major-longon peak, consistent with the termination at $2K$.

The spectra $E$ and $H$ [Eq. (29) of Sec. III A 4] in the lower panels of Fig. 11 (b) show the equipartition tendency at small wavenumbers ($|k| < 10$, say). Such large-scale equipartition can be quantitatively understood as the longon-pulse approximation of the Dirac delta function whose energy spectrum is equipartitioned. [For the zero-Hamiltonian case with uniformly thermalized longons in Fig. 12 to be discussed below, all pulses equally approximate the Dirac deltas, thus the all-scale equipartition of energy.] The nonlocal contribution to $H(|k|)$ at small $|k|$ from $p$ is dominated by small-$|p|$ modes, thus also equipartitioned $H(|k|)$. The scenario similar to that in Fig. 5 (a) and (c) is that the system eventually developes into 'absolute equilibrium' (obvious beyond $t = 40$), with selectively thermalized longons. [Conventional statistical notion can still be used, but with the solitonic longons included.]

The other parts of Fig. 11 present good approximation of and indicate convergence to the GrBH dynamics with a linearly dispersive KdV-type model whose approximate longulence [54] should be acceptable for even more conservative minds, persuasively confirming the appropriateness of the 'longulence' notion.

### 2. Zero-Hamiltonian case

Fig. 12 presents uniformly thermalized longons for the case with the zero-$h$ $u_0$ given by Eq. (28). Again, as seen in Fig. 12 (c), both $b$ and ${}^K g$ typically vanish at the center of each major longon and grow up to maximum

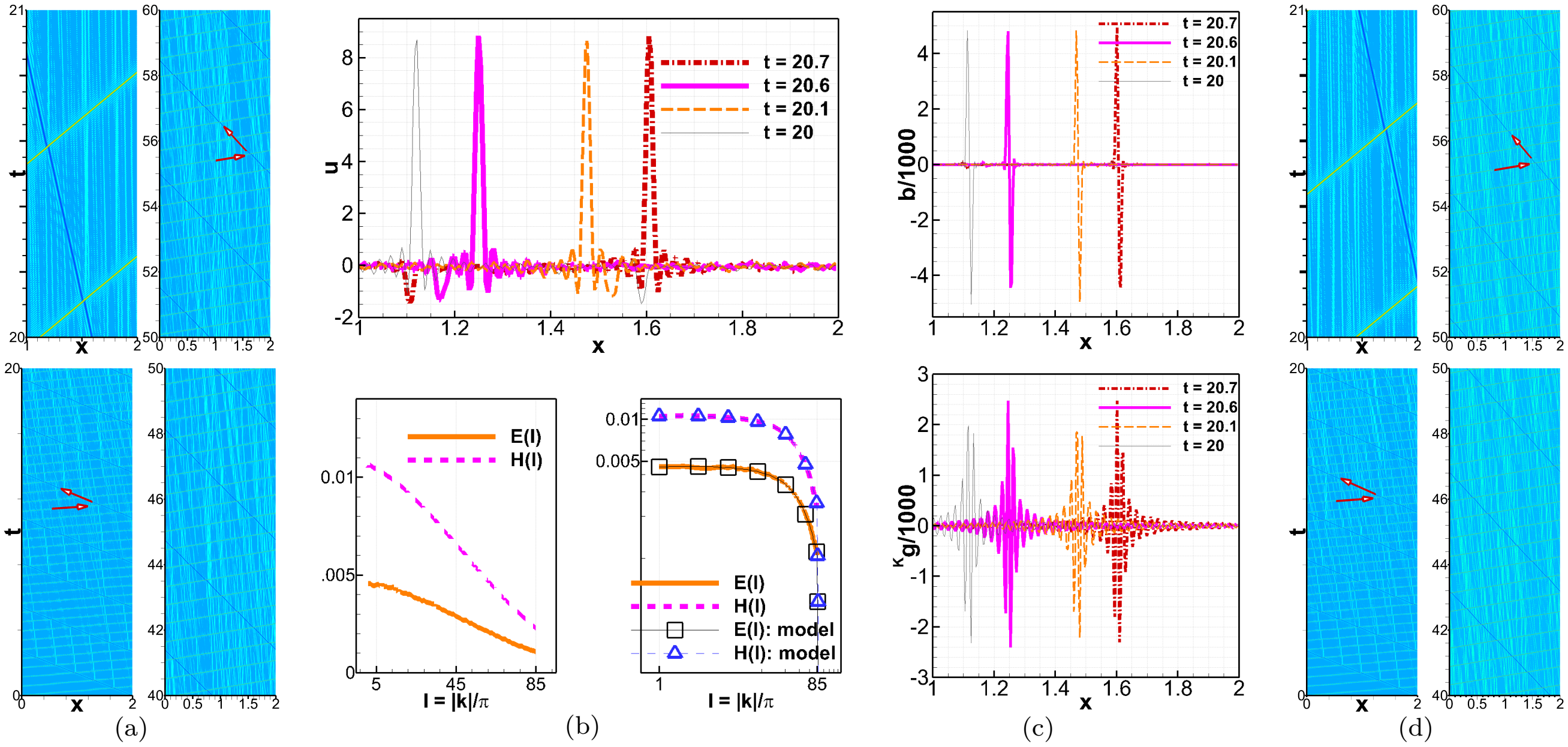

Figure 11. (a) Patterns of the GrBH $u$-contour in different space(horizontal)-time(vertical) regimes, for global view and locally detailed examination. The period is normalized from $2\pi$ to 2. (b) Snapshots of the $u$-profiles (upper frame) corresponding to the region of the upper-left frame in (a) whose color coding can be read from such profiles; and, the energy and Hamiltonian spectra, respectively, $E(l)$ and $H(l)$ with $l = |k|/\pi$, of GrBH (lower-left frame), with their log-log plot (lower-right) compared to those from the approximate hKdV 'model' (thinner lines with sparced symbols to highlight the behaviors within and slightly beyond $K$): the log-log plot highlights the (asymptotic) equipartition in the low wavenumber regime (for $l \leq 5$, say). (c) The nonlinear term $b$ and Galerkin force $^K g$, respectively in the upper and lower frames, at the times corresponding to those of the $u$-snapshots in (b). (d) Space-times patterns of the hKdV model at two regimes for comparison with those of GrBH in (a). The arrows drawn on the solitary trajectories highlight their travelling directions as time goes.

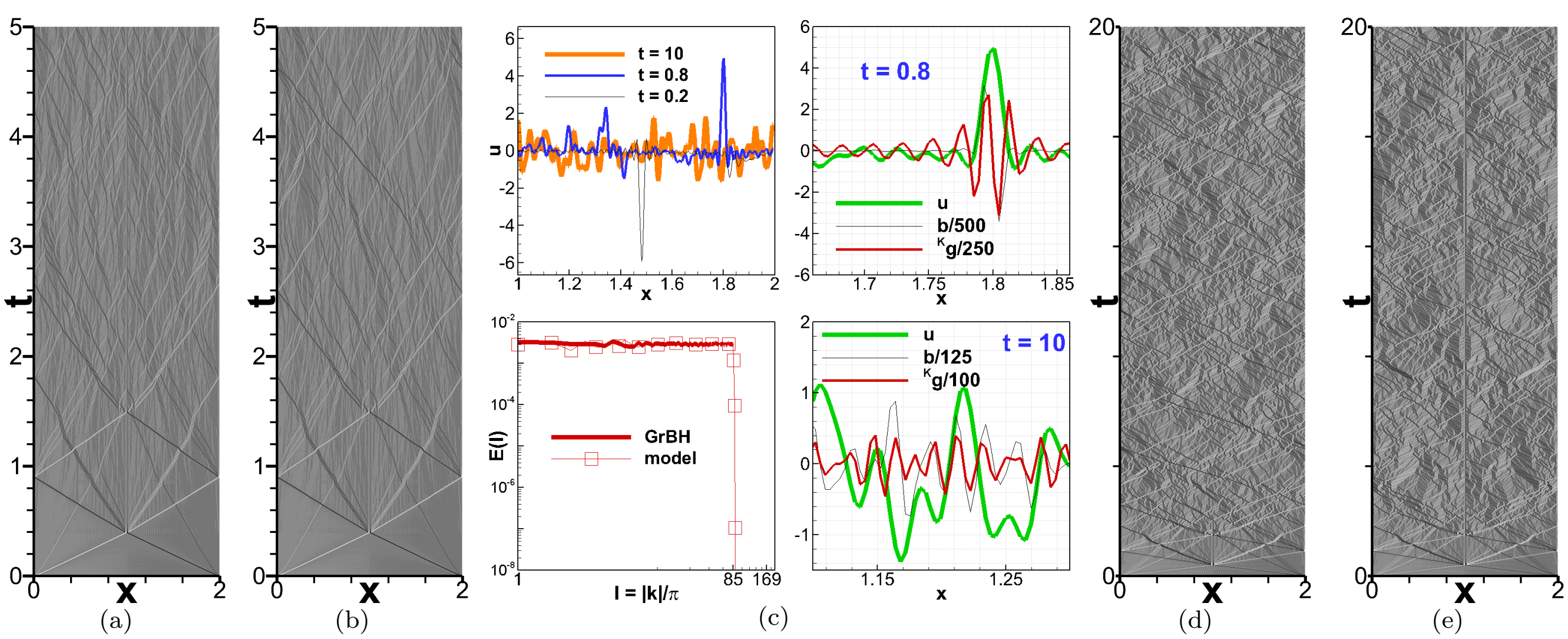

Figure 12. (a) The $u$-contours restricted to the anti-symmetric invariant sub-manifold. (b) The $u$-contours computed without chaos control. (c) The $u$ profiles at some subsequent times (upper-left frame), the energy spectrum of GrBH compared to that of the hKdV 'model' (lower-left) and $u$-$b$-$^K g$ profiles near the locations of one major longon at an earlier time (upper-right) and of several slightly stronger longons at a later time (lower-right). (d) The same $u$-contour of (b), but for longer time. (e) The same $u$-contour of (a), but for longer time.

amplitudes near the HHLs.

Uniformly thermalized cases, such as those in Fig. 5

(e) and Fig. 12, may possess universal properties reflecting the unique GrBH many-mode strong interactions, independent of initial data details or even chaos control. A lesson from classical chaos is that developed longons make sense locally and statistically, and controlling chaos using techniques that restrict numerical solutions to the GrBH (unstable) invariant sub-manifolds offers additional insights into the dynamics. Statistical validity and universality are inherently tied to local accuracy, since the numerical solution at any time can be considered nearly precise, originating from a slightly earlier solution that, due to chaos, is essentially decoupled from much earlier initial data.

For the case in Fig. 12 with $u_0$ anti-symmetric (about $x = \pi$) which is preserved by the GrBH dynamics, the anti-symmetry condition has been incorporated into the computations (Sec. III B) for comparison with the ordinary ones as for other cases without this property. The results show that the overall longon structures are locally and statistically "equivalent" in the following sense: just as the $u$-contours in Fig. 12 [(a) versus (b), and, (d) versus (e)], the computations are accurate for time intervals short but still long enough, with at least a couple of typical longon length (more on this in Sec. IV); and, neither contours present essentially different longons or miss major interaction patterns in the long term.

More examinations comparing the computations with the anti-symmetry condition used up to the developed stages and then relaxed can be found in Appendix B 1 with the additional information about the wavenumber-frequency spectra $\mathtt{E}(k,\omega) := \langle \frac{1}{T^2}\int_{t_0}^{t_0+T}\int_{t_0}^{t_0+T} \hat{u}_k(t)\hat{u}_k^*(t')e^{-\hat{i}\omega(t-t')}dtdt'\rangle$, with the presumably infinite interval $T$ chosen large — over dozeons of the typical longon length — and the time average taken over $t_0$ in the statistically steady regime. Note however that, in general, chaos control can lead to drastically different long-term statistics, i.e., the "climate" (not only the short-time "weather"), as is well-known since von Neumann and Lorenz. For example, stationary longons restricted on the invariant manifold may be obtained (Sec. III C 3 a).

## Appendix B: Longulence in the whole- and sub-manifolds

### 1. The space-time patterns

Numerical experiments corresponding to the zero-Hamiltonian case in Sec. A 2, were performed with the (anti-)symmetry condition employed from the outset, with the initial data of $2\pi$-periodicity, up to times, respectively, $t = 5$ and 20 after which the conditions were turned off for one set of the variables and kept for the other set in the computations. Fig. 13 presents the comparisons of the results, to show the longons are accurate enough 'locally', with the anti-symmetry being clearly maintained for at least one and a half time units, and are meaningful 'statistically' over the long term.

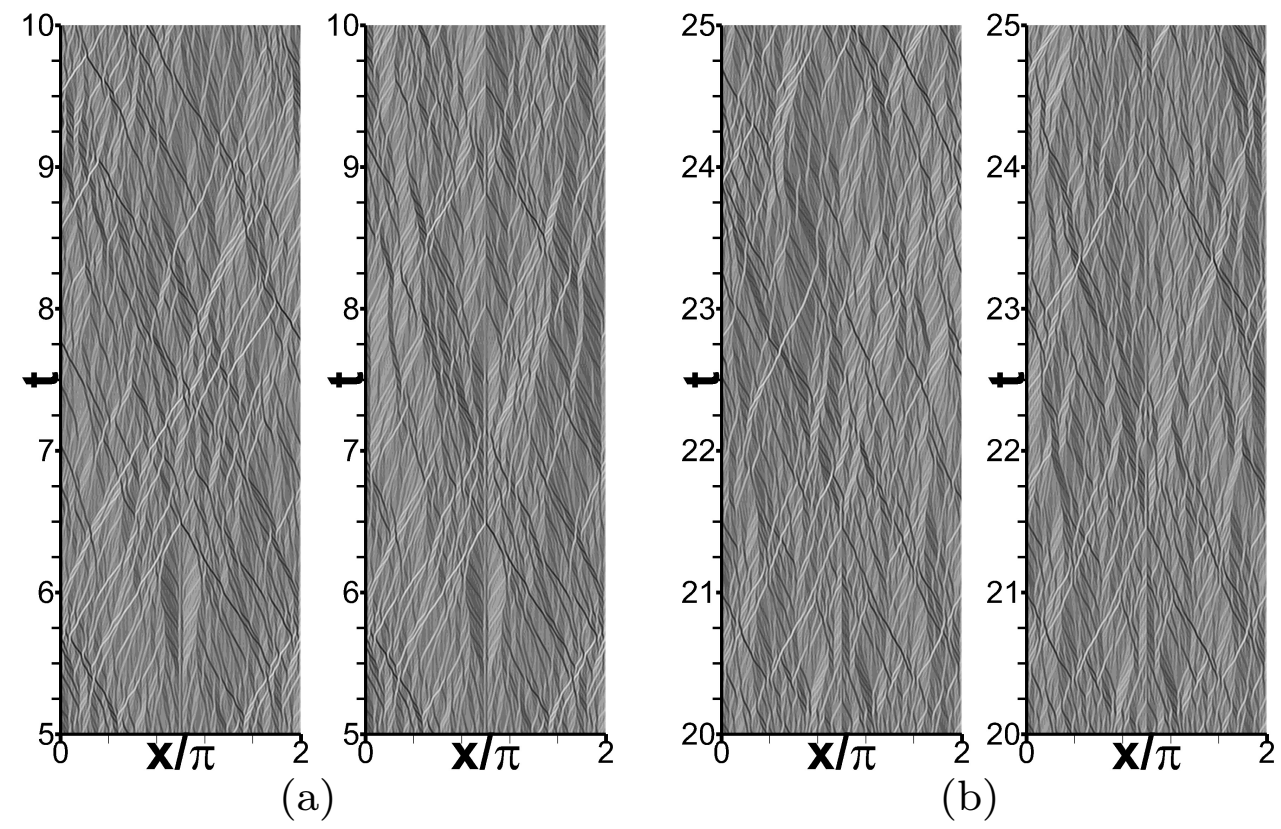

Figure 13. (a) Two $u$-contours from the computations with $u_0 = \sum_{k=1}^{85}\sin(kx)/\sqrt{85}$ continued with (right frame) and without (left frame) the anti-symmetry condition since $t = 5$. (b) Two $u$-contours from the computations with $u_0 = \sum_{k=1}^{85}\sin(kx)/\sqrt{85}$ continued with (right) and without (left) the anti-symmetry condition since $t = 20$.

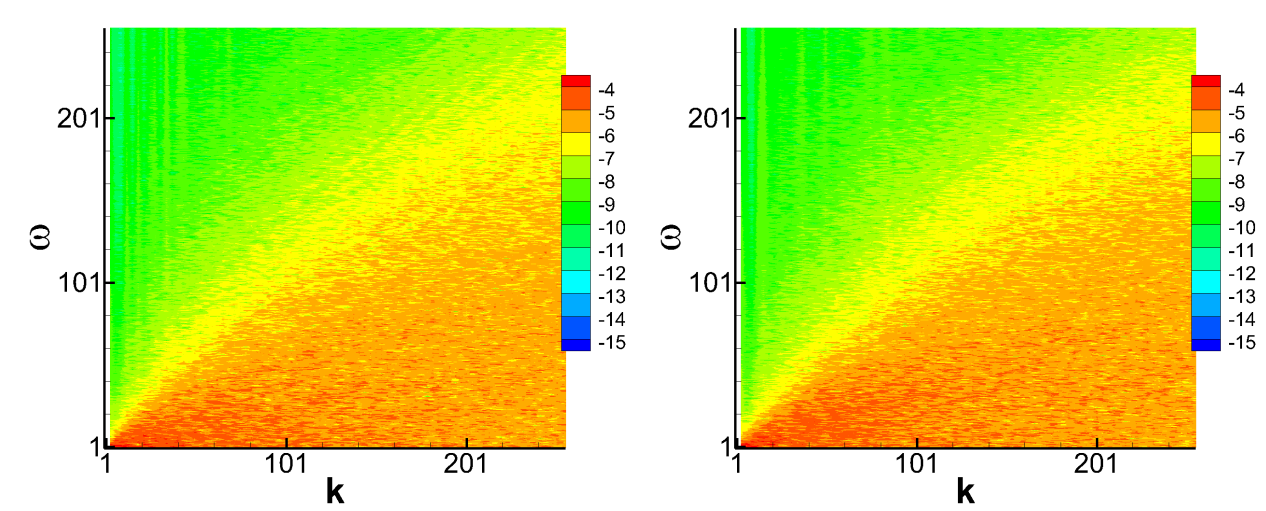

Figure 14. The left and right panels are respectively the logarithmic of the wavenumber-frequency spectra from the fully-developed longulence regimes (after $t = 20$) of the two simulations, (d) and (e) in Fig. 12, without and with using the antisymmetry condition.

### 2. Space-time spectra

Fig. 14 shows nearly the same $\log \mathtt{E}(k,\omega)$ pattern (with sample fluctuations) of developed longons computed with and without the antisymmetry condition (Fig. 12). Such a result confirms that the uncontrolled computations are meaningful.